\documentclass[11pt,
nofootinbib,floatfix]{revtex4}

\usepackage{graphicx}
\usepackage{float}
\usepackage{epsfig}
\newcommand{\be}{\begin{equation}}
\newcommand{\ee}{\end{equation}}
\newcommand{\ba}{\begin{eqnarray}}
\newcommand{\ea}{\end{eqnarray}}

\def\vec#1{{\mbox{\boldmath$#1$}}}

\newcommand{\ep}{\epsilon}

\begin{document}
\begin{titlepage}


\title{
QCD corrections to top quark pair production in association
with a photon  at hadron colliders
}

\author{Kirill Melnikov and Markus Schulze
}
\affiliation{Department of Physics and Astronomy,Johns Hopkins University,\\
Baltimore, MD, USA}
\author{Andreas Scharf}
\affiliation{
Department of Physics,
State University of New York at Buffalo,\\
Buffalo, NY, USA}

\begin{abstract}

\vspace{2mm}

We compute  QCD corrections to the production of
a $t \bar t$  pair in association with a  hard photon
at the Tevatron and the LHC.  This process allows a direct
measurement of the top quark electromagnetic couplings that, at the moment,
are  only loosely constrained.   We include top quark decays,
treating them
in the narrow width approximation, and  retain
spin correlations of final-state particles.
Photon radiation off top quark decay products is included
in our calculation and yields a significant contribution to the
cross-section.
We study next-to-leading order
 QCD corrections to the $p \bar p \to  t \bar t \gamma$ process
at the Tevatron  for the selection criteria used in a recent
measurement by the CDF collaboration.
We also discuss the impact of QCD corrections
to the $pp \to t \bar t \gamma$ process on the measurement  of the top
quark electric charge at the $14$ TeV LHC.

\end{abstract}

\maketitle

\thispagestyle{empty}
\end{titlepage}

\section{Introduction}

More than fifteen years after the discovery of the top quark, many of its
quantum numbers are  still not well-measured  experimentally. For example,
until recently \cite{qtmes} it was  possible to have a consistent description
of ``top'' quark physics, assuming that  the electric charge
of the ``top'' quark is $Q_t = -4/3$, in contrast to its usual value
$Q_t = 2/3$ \cite{ma1}.
By analyzing tracks of  charged hadrons to estimate jet charges,
the D0 collaboration excludes, at less than $2\sigma$ confidence
level, the hypothesis
that  the  event sample  comes {\it entirely} from a
heavy quark with the electric charge $Q_t = -4/3$.
As pointed out in Ref.~\cite{ub},
a more direct way to measure the top quark charge
is to study  the production of a top quark pair in association
with a  hard photon. Indeed,  to an extent that photons are only radiated
off the top quarks, the rate for $t \bar t \gamma$  production
is proportional to the square of  the top quark electric charge.
This assumption works well  at the LHC, once photon radiation
off top quark decay products is suppressed, but it fails at the
Tevatron because top quark pair production there is dominated by
$q \bar q $ annihilation.

The CDF collaboration has recently measured the cross-section
of the process $p \bar p \to t \bar t \gamma$, searching  for
an excess in events that contain a lepton, a photon, $b$-jets and
large missing energy \cite{cdf}. Using
$1.9~{\rm fb}^{-1}$  of data, they observed  nine  events that
they interpreted as due to
the $p \bar p \to t \bar t \gamma$ process.
The analysis will be
improved by using a larger data sample \cite{privatecom}, so
about fifty $t\bar{t}\gamma$ events can be expected in the near future.
It is therefore important to have a reliable prediction for this
process, including the possibility of applying
selection criteria for final state particles
used in Ref.~\cite{cdf}.

The authors of  Refs.~\cite{ub,ub1} analyzed the potential
of the Tevatron and the LHC to study electroweak couplings
of the top quark, focusing on the $t \bar t Z$ and $t \bar t\gamma$ final states.
If we neglect  parity non-conservation,
the interaction of top quarks with on-shell
photons is described by two quantities -- the electric charge $Q_t$
and  the anomalous magnetic moment $a_t$.  Both of these quantities
can be studied in  the $t \bar t \gamma$
production process.  As shown in
Refs.~\cite{ub,ub1}, the best sensitivity
to $Q_t$ at the LHC is obtained if  kinematic cuts force the photon to
be emitted either in the  production stage
or in the decay stage. Because the
non-vanishing anomalous magnetic
moment of the top quark corresponds to a
dimension-five non-renormalizable operator, it  leads to
a harder spectrum of photons in $pp (p \bar p) \to t \bar t \gamma$.
Given enough statistics, it should be possible to study
this effect experimentally.   Regardless of the details
pertinent to a particular measurement,
once selection criteria are specified, the study  of
top quark electromagnetic couplings
becomes  a  counting experiment which
may be subject to significant higher
order QCD corrections. Therefore, the   computation of
next-to-leading order (NLO) QCD corrections to  $t \bar t \gamma$
hadroproduction,
that correctly incorporates
   decays of top quarks,  becomes important.

NLO QCD corrections to the production of
a $t \bar t$  pair and a hard photon in hadron collisions
were
recently calculated   by Duan {\it et al.}~\cite{ma}. This  computation
was performed in the approximation of  stable top quarks. While
such an approximation gives an idea  about
the significance  of higher order QCD effects for the
production of $ t \bar t \gamma$, it can not be used to find
the magnitude of NLO QCD corrections when specific cuts are imposed
on  top quark decay products.  As we explained in the
previous paragraph, the ability to do this is important  for a
realistic analysis.
In this paper, we extend the results  of
Ref.~\cite{ma} by
computing NLO QCD corrections to  $ pp(p \bar p) \to t \bar t + \gamma$,
allowing for  decays
of  top quarks.  We note that radiative decays of top quarks
are included into our analysis.
To calculate one-loop virtual amplitudes,
we employ the method  of generalized $D$-dimensional unitarity
suggested in Ref.~\cite{Giele:2008ve} and extended  to massive
particles in Ref.~\cite{egkm}.  The current
paper builds upon the  previous
studies of $t \bar t$
and $t \bar t+j$  production in hadron collisions, performed by two
of us \cite{ms,msj}.
Many technical aspects of the calculation are explained in those
references and we do not repeat them here.

When top quarks are treated as truly unstable particles,
non-factorizable QCD corrections appear \cite{my}.
Non-factorizable corrections imply a cross-talk
between production and decays of top quarks; they can not
be described in the narrow width approximation.  It is
well-understood by now \cite{my} that, in many cases, these non-factorizable
corrections lead to effects
that are suppressed  by  ${\cal O}(\alpha_s \Gamma_t /m_t)$,
instead of the naive  ${\cal O}(\alpha_s)$ expectation for the suppression.
Recently, the smallness of non-factorizable corrections  in reactions
with top quarks  was confirmed by an explicit computation
of the NLO QCD corrections to $pp(p \bar p) \to W^+W^-b \bar b$ process
that included both factorizable and non-factorizable contributions
\cite{Denner:2010jp,Bevilacqua:2010qb}.
In what follows we ignore the
non-factorizable corrections and
work in the on-shell approximation for top quarks.

As a final comment, we note that NLO QCD corrections are known for
two other processes where the top quark pair is produced in association
with color-neutral objects --
$pp \to t \bar t H$  \cite{Beenakker:2002nc, Dawson:2003zu}
and  $pp \to t \bar t Z$ \cite{Lazopoulos:2008de} ---
and to the production of a $t \bar t$ pair in association with
one \cite{Dittmaier:2007wz,Dittmaier:2008uj,Bevilacqua:2010ve,msj}
and two \cite{Bevilacqua:2010ve} jets,
as well as in association with a $b \bar{b}$ pair \cite{Bredenstein:2008zb,Bredenstein:2009aj,Bevilacqua:2009zn,Bredenstein:2010rs}.
In all the cases, the NLO QCD corrections
are calculated  either assuming that all final state particles are stable,
or treating QCD radiation in top decays incompletely\footnote{
See however Ref.~\cite{Kardos:2011qa} where QCD radiation in top quark decays is included in the computation of the top quark pair production cross-section
in association with one jet by means of a parton shower.}.
Similarly to the $t \bar t \gamma$ case, removing these omissions  may
become important 
for precision phenomenology, especially
when aggressive cuts are involved to separate signals from backgrounds.

The remainder of the paper is organized as follows. In Section~\ref{sect2}
we  describe the setup of the calculation
and present some results for the case when top quarks are treated
as stable particles. In Section~\ref{sect3} we discuss the
computation of NLO QCD
corrections to radiative decays of top quarks. In Section~\ref{sect4}
we present phenomenological studies relevant for the
Tevatron and the LHC, including decays of top quarks.
We conclude in Section~\ref{sectc}. Technical details of the calculation
are described in the Appendices.

\section{Production of a $t \bar t $ pair
and a photon: stable top quarks}
\label{sect2}

We first discuss the case of stable top quarks.
To compute the NLO QCD corrections to $pp( p \bar p)  \to t \bar t \gamma$,
we need to calculate one-loop virtual corrections and to account for the
emission of an additional  massless parton.
For the calculation of  the virtual corrections, we employ
the method  of generalized $D$-dimensional unitarity
suggested in Ref.~\cite{Giele:2008ve}.  This
method has been used earlier by two of us
in  the computation of hadroproduction of $t \bar t+{\rm  jet}$ in
Ref.~\cite{msj}.  To describe the $t \bar t \gamma$ final state, we can re-use
much of that computation. For example, linear combinations
of color-ordered one-loop amplitudes for $0 \to t \bar t +3g$ \cite{msj}
give color-ordered amplitudes for $0 \to t \bar t + 2g + \gamma$ \cite{bdk}.
Similarly, color-ordered
amplitudes for $0 \to t \bar t + q \bar q + g$ \cite{msj}
can be  used to construct color-ordered amplitudes for
$0 \to t \bar t + q \bar q + \gamma$.   All the details of how
amplitudes with gluons and quarks are transformed into amplitudes with
gluons, quarks and a photon are given in Appendix~\ref{a1}.
We have checked our results
for virtual corrections by re-calculating them, for a few phase-space
points by using  an independent implementation of the Ossola-Pittau-Papadopoulos
(OPP) procedure \cite{opp}, that we apply to individual Feynman diagrams.
The Feynman diagrams are generated with the package FeynArts \cite{Hahn:2000kx}.

The second, logically distinct part of any one-loop computation is the
calculation of  real emission corrections. When integrated
over available phase-space, these corrections diverge.
Such divergences must be
removed by an appropriate  procedure. We use the dipole formalism of
Ref.~\cite{Catani:1996vz} extended  to deal with
QCD radiation off  massive particles in Ref.~\cite{Catani:2002hc}.
Dipoles  relevant for our calculation can be found in
Refs.~\cite{Bevilacqua:2009zn,mcfm,Campbell:2005bb}.
In the actual implementation of the subtraction procedure, we closely
follow Ref.~\cite{mcfm}.  We have checked that our results do not
depend on the parameter  that restricts the integration over the dipole
phase-space; this is a useful way to control the consistency of the
implementation of the subtraction terms and to improve the
efficiency of the computation \cite{nagy}.

As we mentioned earlier, the calculation
of NLO QCD corrections to $t \bar t \gamma$ production in hadronic
collisions
for stable top quarks
was reported in Ref.~\cite{ma}.  When we choose  the setup
of the calculation as close to Ref.~\cite{ma} as possible,
we get good agreement with their results. However, some
choices made in Ref.~\cite{ma} -- for example the use of the electromagnetic
coupling at the scale $M_Z$,
the use of charm and bottom masses in the computation of the partonic channels
$q g \to t \bar t \gamma +g$, $q = c,b$  and the three degree
cut  on the opening angle between the photon and the light quark in the
final state  -- do not look very appealing to us.  For this reason, we
decided to present a number of results for cross-sections and kinematic
distributions which can not be directly compared with the results
reported in Ref.~\cite{ma} but which, we believe, correspond to  more
realistic choices of input parameters and better resemble details
of experimental analyses.

\begin{figure}[!t]
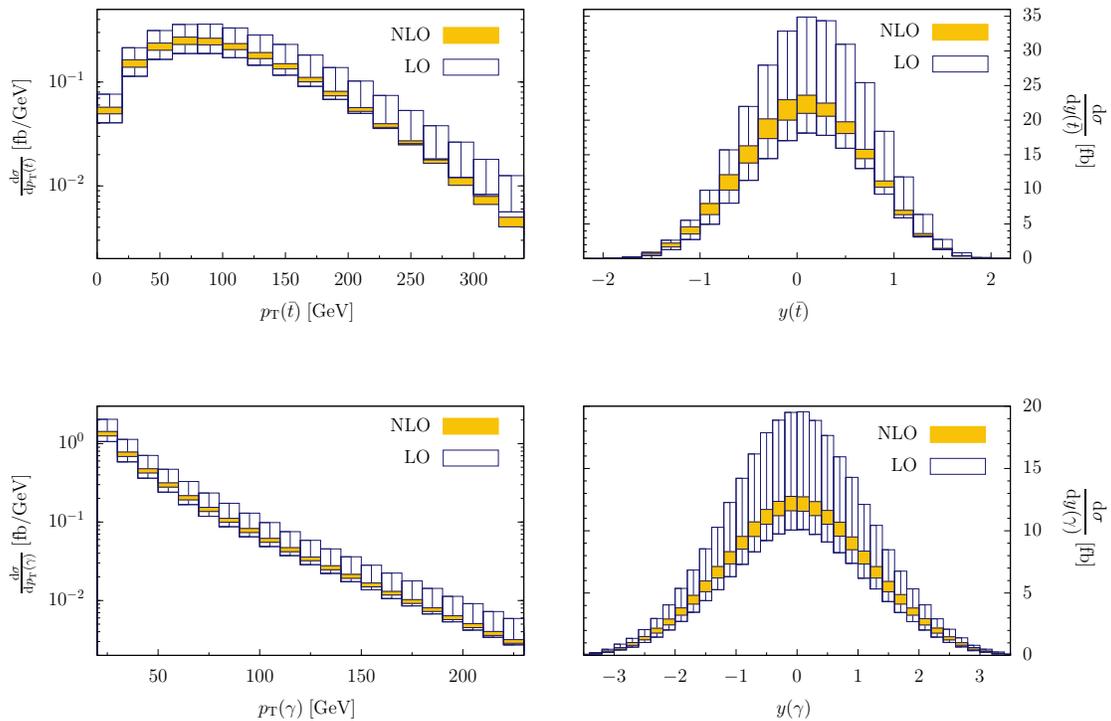

 \begin{center}
  \scalebox{0.5}{\input{TEV_20_Fig01.tex}}
  \scalebox{0.5}{\input{TEV_20_Fig02.tex}} \\[8mm]
  \scalebox{0.5}{\input{TEV_20_Fig03.tex}}
  \scalebox{0.5}{\input{TEV_20_Fig04.tex}} \\[8mm]
\end{center}
\caption{ Kinematic distributions
in the production of a  $t \bar t$ pair and a hard photon
at the Tevatron, for stable top quarks.
The bands correspond to the variation of
the renormalization and factorization scales
in the interval $m_t/2 < \mu <  2m_t$.
Upper (lower) panes show $\bar t\;$  ($\gamma$) transverse momentum and
rapidity distributions, respectively.
}
\label{fig1}
\end{figure}

We now turn to the discussion of our  results for the hadroproduction
of a $t \bar t$ pair and a hard photon, for stable top quarks.
Throughout the paper,
we choose the top quark mass $m_t = 172~{\rm GeV}$
and parton distribution
functions CTEQ6L1 and CTEQ6.6M
\cite{Pumplin:2002vw,Nadolsky:2008zw}
for leading and next-to-leading computations,
respectively.
The strong coupling constant $\alpha_s$ is evaluated using
one- and two-loop running with five massless flavors.
To describe emission of the real photon, we use the
fine structure constant $\alpha_{\rm QED} = 1/137$.
Although this choice should be self-evident because of QED Ward identities,
we emphasize this fact because in many
previous studies of the  $t \bar t \gamma$ production,
the cross-section was computed with
$\alpha_{t \bar t \gamma} = \alpha_{\rm QED}(M_Z) = 1/128$.
Using the correct value of the fine structure constant
is numerically important because it decreases the prediction for
the $t \bar t \gamma$ cross-section
by about six percent.

For both the Tevatron and the LHC, we require that the photon is relatively
hard  $p_{\perp,\gamma} > 20~{\rm GeV}$ and that it is isolated.
To ensure that the implementation of photon isolation  does not
violate infra-red and
collinear safety, we employ the procedure described in
Ref.~\cite{Frixione:1998jh}. The photon is {\it not} considered
isolated and  events are rejected unless
the  condition
\be
\sum \limits_{i \in {\rm partons}}^{} E_{\perp,i} \;
\theta \left (R - R_{i\gamma} \right )  \le
E_{\perp, \gamma} \left (
\frac{ 1 - \cos(R) }{1 - \cos (R_{\gamma j})} \right )
\label{eq2}
\ee
is fulfilled
for cones of sizes $R$ that are smaller than
$R_{\gamma j} = 0.4$.  In  Eq.(\ref{eq2})  $R_{\gamma i}$ is
the photon-parton
angular distance
$ R_{\gamma i} =\sqrt{(y_\gamma - y_i)^2 + (\varphi_\gamma - \varphi_i)^2}$,
where $y_{\gamma, i}$ ($\varphi_{\gamma, i})$  are the laboratory frame
rapidities
(azimuthal angles)
of the photon and the  parton $i$, respectively.
Also,
$E_{\perp, i}$ is the transverse energy of the parton $i$ and
$E_{\perp, \gamma}$ is the transverse energy of the photon.
We apply all other selection
criteria to jets if and only if their separation from  a photon
exceeds $R_{\gamma j}$. A jet reconstructed inside the
cone of size $R_{\gamma j}$ is not subject to selection criteria,
see Ref.~\cite{Frixione:1998jh}.
As our default, we set the  renormalization and factorization scales
equal to each other  and choose them to be equal to the
mass of the top quark $\mu = m_t$.
We find the cross-section for $p \bar p \to t \bar t \gamma$
at the Tevatron ($\sqrt{s} = 1.96~{\rm TeV}$) to be
\be
\sigma_{\rm LO} = 39.97^{+16.77}_{-10.91}~{\rm fb},
\;\;\; \sigma_{\rm NLO} = 37.6^{+0.8}_{-3.7}~{\rm fb},
\label{eq22}
\ee
where the lower value correspond to the scale set to $\mu = 2 m_t$ and the
upper value to the scale set to $\mu = m_t/2$.  QCD corrections greatly
reduce the uncertainty in the predictions for the cross-section, changing
it from about thirty percent at leading order
to about  ten percent at next-to-leading order.
The cross-section for $pp \to t \bar t \gamma$ at the $14~{\rm TeV}$ LHC
is
\be
\sigma_{\rm LO} = 1.96^{+0.64}_{-0.45}~{\rm pb},
\;\;\;\sigma_{\rm NLO} = 2.93^{+0.42}_{-0.39}~{\rm pb}.
\label{eq33}
\ee
The residual scale uncertainty in  the NLO QCD
cross-section  at the LHC
is about fifteen   percent.

We note that  results  for  $t \bar t \gamma$
production in Eqs.(\ref{eq22},\ref{eq33})
show  significant differences in the QCD corrections
at the Tevatron and the LHC. At the scale $\mu=m_t$,  the
NLO QCD corrections decrease the $t \bar t \gamma$
cross-section by about six percent
at the Tevatron and increase the $t \bar t \gamma$
cross-section by about $55\%$ at the
LHC.
It is peculiar that the magnitude of the NLO QCD corrections to
$t \bar t \gamma$ production at the Tevatron and the LHC is
{\it very similar}
to the magnitude of the NLO QCD corrections  to $t \bar t$ pair production.
While the degree of the correlation between these corrections
is perhaps somewhat surprising,
it can be understood, at least partially, by considering  emissions
of {\it soft} photons which must factorize from the production process
even when the NLO QCD corrections are included.  This may also
explain why a very similar pattern of QCD corrections
was reported in Ref.~\cite{ma},
despite the fact that somewhat different input parameters are used
in that computation.

\begin{figure}[!t]
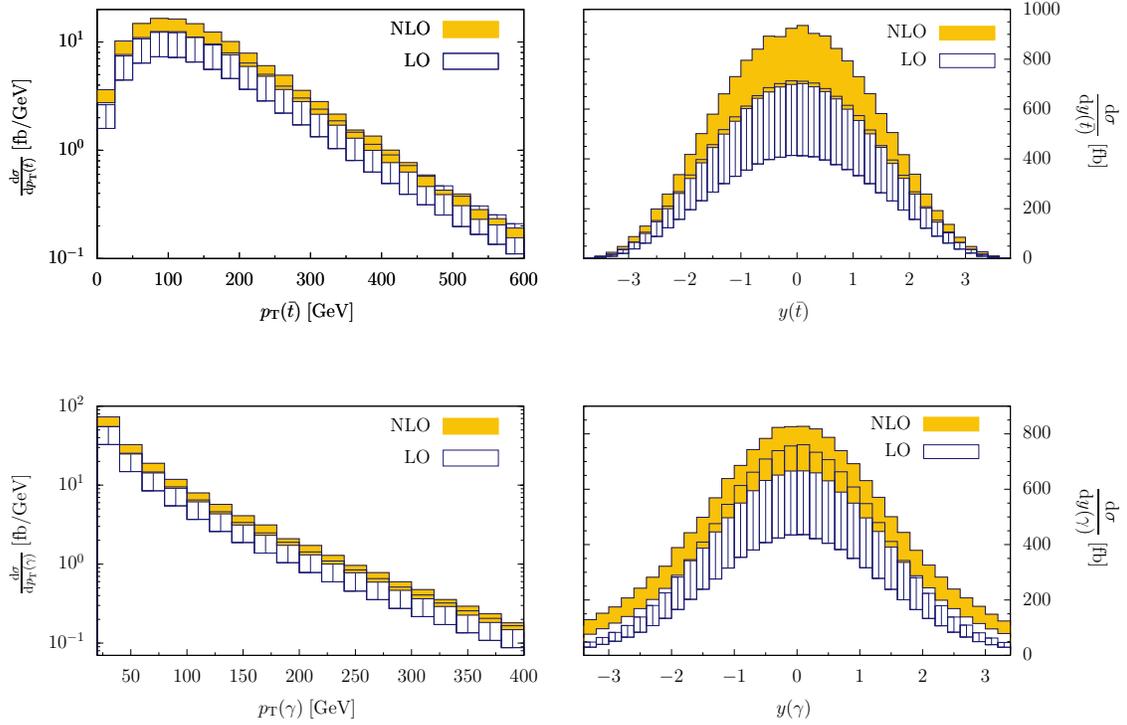

 \begin{center}
  \scalebox{0.5}{\input{LHC_21_Fig01.tex}}
  \scalebox{0.5}{\input{LHC_21_Fig02.tex}} \\[8mm]
  \scalebox{0.5}{\input{LHC_21_Fig03.tex}}
  \scalebox{0.5}{\input{LHC_21_Fig04.tex}} \\[8mm]
\end{center}
\caption{
Kinematic distributions
in the production of a $t \bar t$ pair and a hard photon
at the LHC, for stable top quarks.
The bands correspond to the variation of
the renormalization and factorization scales
in the interval $m_t/2 < \mu <  2m_t$.
Upper (lower) panes show $\bar t\;$  ($\gamma$) transverse momentum and
rapidity distributions, respectively.
}
\label{fig2}
\end{figure}

Kinematic distributions of
top quark and photon transverse momenta and rapidities
at  the   Tevatron and the LHC, are shown
in Figs.~\ref{fig1},\ref{fig2}.  No dramatic changes in shapes
of these distributions are observed.
However, the rapidity distribution of top quarks at the Tevatron
exhibits some interesting features.  Indeed, in the process
$p \bar p \to t \bar t \gamma$, photons can be emitted from
both  initial and final states. Interference of these emissions
gives rise to a charge or forward-backward asymmetry of top quarks.
 We define
\be
A_{t} = \frac{\sigma(y_{t} > 0)
- \sigma(y_{t} < 0)}{\sigma(y_{t}> 0)
+ \sigma(y_{t} < 0)},
\ee
where $\displaystyle
y_t = \frac{1}{2} \ln\left( \frac{E_t+p_{t,z}}{E_t - p_{t,z}} \right)
$
is the rapidity of the top quark in the laboratory frame.
As was pointed out in Ref.~\cite{ma}, the forward-backward
 asymmetry is significant.
Calculating it in leading and next-to-leading order
in perturbative QCD, with the parameters specified at the beginning
of this  Section, we obtain
\be
A_{t,\rm LO} = -17.2^{+0.0}_{-0.0}\,\%,\;\;\;\;\;
A_{t, \rm NLO} = -11.9^{+2.9}_{-1.3}\, \% ,
\label{eq222}
\ee
where the central value corresponds to the renormalization and factorization
scales set to $\mu = m_t$ and the lower(upper) value to
$\mu = m_t/2$ and $\mu =2 m_t$, respectively.

It is peculiar  that the change in
the NLO QCD asymmetry  is {\it nearly identical} to the
size of the NLO QCD corrections to  forward-backward asymmetry
in $p \bar p \to t \bar t j$,  computed
in Refs.~\cite{Dittmaier:2007wz,Dittmaier:2008uj,msj}. In fact, this  similarity
of NLO QCD corrections to the asymmetries
in $ p \bar p \to t \bar t \gamma$ and $p \bar p \to t \bar t j$ is
easy to understand, following the observation of  Ref.~\cite{msj} that
large NLO QCD correction  to  the asymmetry in $p \bar p \to t \bar t j$
is  related
to  the $5\%$ asymmetry  in the {\it inclusive}
rate for $p \bar p \to t \bar t$,  first computed in Ref.~\cite{kuhn}.

\section{ QCD corrections to radiative decays of top quarks
}
\label{sect3}

In this Section,  we describe the  computation of  the NLO QCD corrections
to the radiative decay of the top quark $t \to b W \gamma$.
Since this is a relatively low-multiplicity process,
the calculation of virtual corrections
is performed using conventional Feynman diagrams,
generated with FeynArts \cite{Hahn:2000kx},
and the Passarino-Veltman reduction \cite{pv}. For a few kinematic points,
the results are checked against a computation based upon an
independent implementation of the OPP procedure \cite{opp},
that is applied to individual Feynman diagrams.

The real emission corrections to the decay rate
are computed using the subtraction formalism described
in Ref.~\cite{ekt}, in the context of non-radiative top
decay $ t \to b W$.  However, the subtraction is also applicable
to the radiative decay $t \to b W \gamma$  if one replaces
the invariant mass of the $W$-boson with the invariant mass
of the $W$-boson and the photon
in all the relevant formulas in Ref.~\cite{ekt}. Specifically,
the subtraction term, required to make the real emission contribution
$ t \to W b \gamma + g$ integrable is given by the product of the matrix element
squared for the process $ t \to W b \gamma$ and the dipole that
reads \cite{ekt}
\be
D(p_t,p_g,p_b) = 4\pi \alpha_s \mu^{2\epsilon} C_F
\left [ \frac{1}{p_b p_g} \left ( \frac{2}{1-z} - 1-z
- y \epsilon (1-z) \right ) - \frac{m_t^2}{(p_t p_g)^2}
\right ].
\label{eq5}
\ee
The kinematic variables used in Eq.(\ref{eq5}) are defined through
\be
p_b p_g = \frac{m_t^2}{2} (1 - r)^2 y,\;\;\;\;\;
p_t p_g = \frac{m_t^2}{2} ( 1- r^2)(1-z),
\ee
where  $r^2 = (p_W + p_\gamma)^2/m_t^2$.  In Ref.~\cite{ekt},
the dipole in Eq.(\ref{eq5})
is integrated over the unresolved phase-space
\be
\int \left [{\rm d}g \right ] \; D(p_t,p_g,p_b)
=
{\cal N} \int \limits_{0}^{1}
{\rm d} z \left ( r^2 + z(1-r^2) \right )^{-\ep}
\int \limits_{0}^{y_{\rm max}}
{\rm d} y y^{-\ep} (y_{\rm max} - y)^{-\ep}
D(p_t,p_g,p_b),
\label{eq6}
\ee
where $\ep = (4-d)/2$ is the parameter of the dimensional
regularization, $d$ is the number of space-time dimensions and
\be
y_{\rm max} = \frac{(1+r)^2z(1-z)}{z+r^2(1-z)},\;\;\;\;\;\;\;
{\cal N} = \frac{(1-r)^2}{16\pi^2} m_t^{2-2\ep}
\frac{(4\pi)^\ep}{\Gamma(1-\ep)}
\left (\frac{1+r}{1-r}   \right )^{2\ep}.
\ee

It is convenient to
restrict the subtraction counter-terms to parts of the phase-space
that are not too far from the singular limits \cite{nagy}.
If this is done,
the subtraction terms need to be modified.
Introducing such a modification in Eq.(6) and integrating
over restricted phase-space, we find
\ba
&& \int \left [ {\rm d}g \right ] \; D(p_t,p_g,p_b)
\left [ 1-\theta(1-\alpha-z) \theta(y- \alpha y_{\rm max} ) \right ] =
\nonumber \\
&& \frac{\alpha_s C_F}{2\pi}
\frac{(4\pi\mu^2)^\epsilon}{m_t^{2\ep} \Gamma(1-\ep)}
\left [ \frac{1}{\ep^2}
+\frac{1}{\ep} \left ( \frac{5}{2} - 2\ln(1-r^2) \right )
+ \frac{25}{4}
+ \frac{1}{2}
\left (
\frac{1}{(1-r^2)^2} - \frac{8}{(1-r^2)} + 7 \right ) \ln r^2
\right.
\nonumber \\
&&  \left.
+ \frac{1}{2(1-r^2)}
+ 2 {\rm Li}_2(1-r^2) - \frac{5\pi^2}{6} - 5 \ln(1-r^2) +
2 \ln^2(1-r^2) + \frac{\eta}{2}
\right.
\nonumber \\
&& \left.
- 2 \ln^2 \alpha
- \left ( \frac{7}{2} - 4\alpha
+ \frac{\alpha^2}{2} \right ) \ln \alpha
+\frac{2(1-\alpha)r^2}{1-r^2}
\ln \left ( \frac{r^2}{1-\alpha+r^2\alpha} \right )
\right ].
\ea

\vspace*{0.5cm}
We now present some numerical results for the QCD corrections to the radiative
decay of the top quark $t \to b W \gamma$. We use
$\alpha_s(m_t) = 0.107691$, which corresponds to the CTEQ NLO value
of the strong coupling constant at $\mu = M_Z$ supplemented
with the two-loop running to $\mu = m_t$.
We take the mass of the $W$-boson to be
$M_W = 80.419~{\rm GeV}$.  We work in the top quark
rest frame and require the photon energy $E_\gamma$  to be
larger than $ 10~{\rm GeV}$ and the opening angle
between the momentum of the bottom quark and the photon to
be such that  $\cos \theta_{b \gamma} < 0.98$.
With these input parameters, we obtain  the radiative
decay width of the top quark $ t \to b W \gamma$ at leading and
next-to-leading order in perturbative QCD
\be
\Gamma_{\rm LO} = 4.48~{\rm MeV},\;\;\;\;\Gamma_{\rm NLO} = 3.89~{\rm MeV}.
\ee

For the choice of the kinematic cuts described above,
the QCD radiative corrections decrease radiative decay width by thirteen
percent. This is similar to, but somewhat larger than,
the magnitude of the NLO QCD corrections
to the top quark decay width $ t \to b W$, which decrease the
decay width $t \to Wb$ by about eight percent.

\begin{figure}[t]
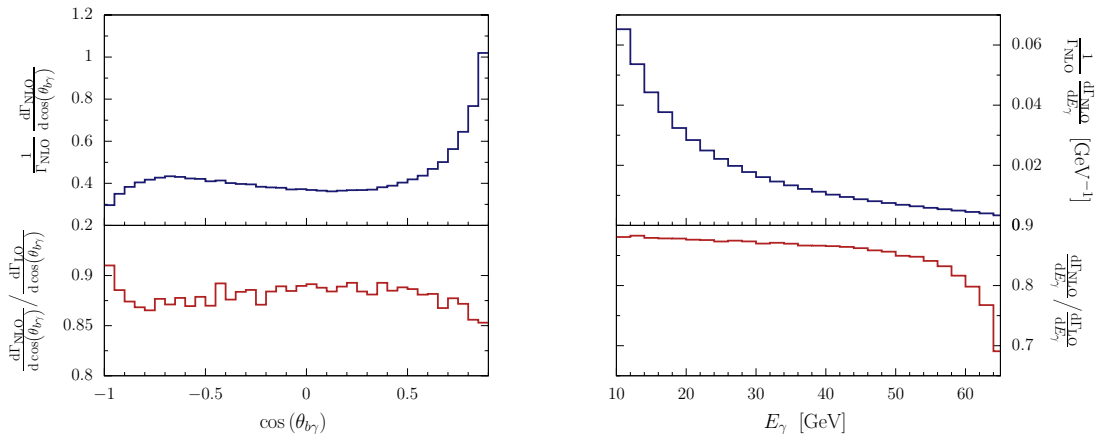

 \begin{center}
\hspace*{0.5cm}
\scalebox{0.45}{\input{dGamma_Angle.tex}}
\hspace*{-2cm}
\scalebox{0.45}{\input{dGamma_Egamma.tex}} \\[8mm]
\end{center}
\caption{Kinematic distributions in  radiative decays
of top quarks $t \to W b \gamma$.
Upper panes
show normalized distributions of the cosine of the angle between
the $b$-quark and the photon and of the photon energy,
computed through NLO QCD.
Lower panes show ratios of NLO and  LO kinematic distributions.
}
\label{fig3a}
\end{figure}

In Fig.~\ref{fig3a},
we show distributions of
the opening angle between the bottom quark and
the photon and
of  the photon energy,  and ratios of NLO and LO distributions.
Shapes of these distributions
are perfectly described by leading order computations;
the NLO QCD corrections provide an overall renormalization
factor.  The distribution of the photon
energy shows  canonical
enhancement of the soft photon emission probability at low
$E_\gamma$, while the distribution in the opening
angle shows a collinear enhancement peak at small relative angles between
the bottom quark and the photon. Suppressing emissions from
bottom quarks is important for the analysis of the top quark charge
that we discuss in the next Section;  a simple  way to accomplish
this  is to require that the $b$-jet and the hard photon are
sufficiently separated.

\newpage

\section{
Production of a $t \bar t $ pair
and a photon: unstable  top quarks}
\label{sect4}

A realistic description of the hadroproduction of  the $t \bar t \gamma $
final state
requires including decays of top quarks.  This is important for two
reasons: first, it defines realistic acceptances  and, second, photons can be
radiated from the top quark decay products and it may  be important to take this
effect into account.  In this Section, we present some results for
realistic selection cuts at the Tevatron and the LHC.

The computation  described
below is  performed within the following framework.
For both the Tevatron and the LHC, we consider the lepton plus jets
channel
$pp, (p \bar p) \to
(t \to W^+(l^+\nu) b  )
(\bar t \to W^- (jj) \bar b ) \gamma $.
Top quarks are treated in the narrow width approximation and
all spin correlations are retained.
We include decays of $W$-bosons
into leptons of definite flavor
($e$ {\it or} $\mu$ but not both) and hadronic decays of $W$-bosons
into two families of light quarks, that are always treated
as massless.
The
$W$-bosons are on their mass-shells and we do not consider QCD radiative
corrections to hadronic decays of $W$-bosons.
We include photon radiation
in the production of a $t \bar t$ pair
and photon radiation in the decays of top quarks. We note that
photons
can be radiated by any charged particle in the top quark decay
process, including  the decay products of  $W$-bosons.

Before describing the results of the computation, we summarize the
input parameters.
For numerical calculations we use  $m_t = 172~{\rm GeV}$,
$M_W = 80.419~{\rm GeV}$,
the value of the Fermi constant
$G_\mathrm{F} = 1.16639 \times 10^{-5}~{\rm GeV}^{-2}$ and
$\Gamma_W = 2.14~{\rm GeV}$.
We use CTEQ6L1 and CTEQ6.6M
\cite{Pumplin:2002vw,Nadolsky:2008zw} parton distribution
functions, for leading and next-to-leading computations,
respectively. For $\alpha_s$ we use one(two)-loop running
for leading(next-to-leading) order calculations, neglecting
the contribution of top quarks to the QCD $\beta$-function.
For reference, we give numerical values  for the top quark
width $\Gamma(t \to Wb)$ at leading and next-to-leading
order \cite{kj}
\be
\Gamma_t^{\rm LO} =
\frac{G_\mathrm{F} m_t^3}{8 \sqrt{2} \pi} \left ( 1 - \frac{M_W^2}{m_t^2}
\right )^2 \left ( 1 + 2 \frac{M_W^2}{m_t^2} \right )
= 1.4653~{\rm GeV},\;\;\;
\Gamma_t^{\rm NLO} = 1.3396~{\rm GeV},
\ee
where $\Gamma_t^{\rm NLO}$ is  calculated with $\alpha_s(m_t) = 0.107691$.

\begin{figure}[!t]
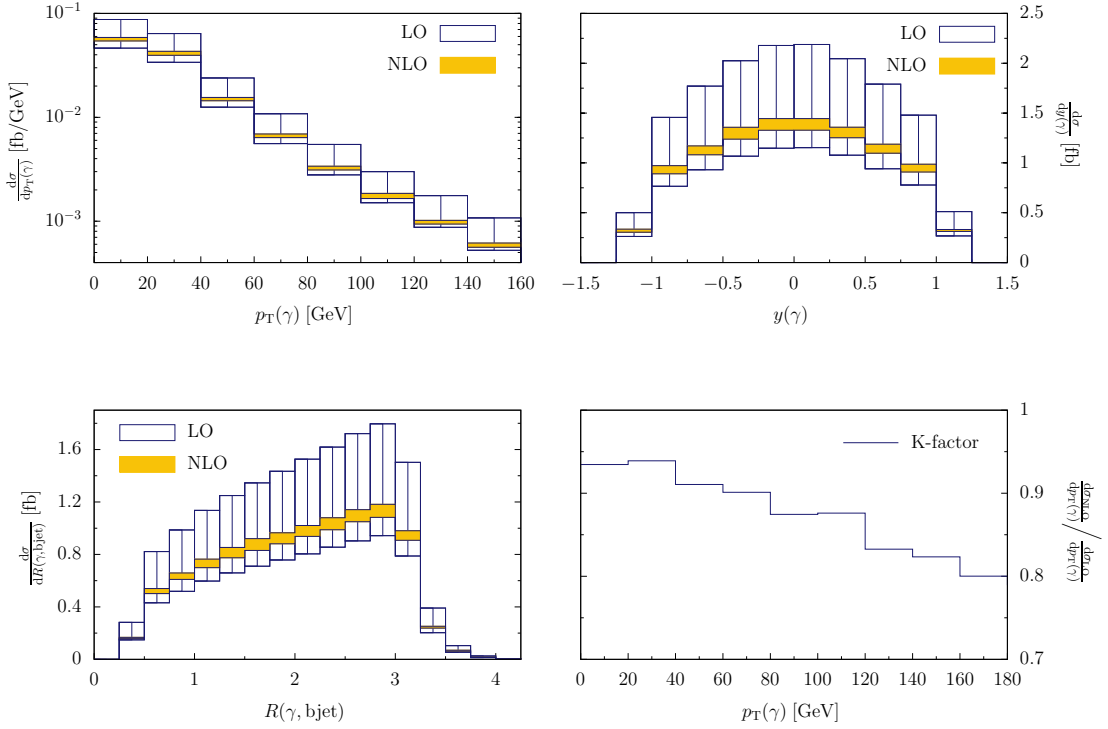

 \begin{center}
  \scalebox{0.5}{\input{TEV_24_Fig01.tex}}
  \scalebox{0.5}{\input{TEV_24_Fig02.tex}} \\[8mm]
  \scalebox{0.5}{\input{TEV_24_Fig07.tex}}
  \scalebox{0.5}{\input{TEV_24_Fig11.tex}}
\end{center}
\caption{
Kinematic distributions of photons
in $p \bar p \to
(t \to W^+(l^+\nu) b  )
(\bar t \to W^- (jj) \bar b ) \gamma $ process
at the Tevatron. The bands correspond to the variation of
the renormalization and factorization scales
in the interval $m_t/2 < \mu <  2m_t$. We show the
transverse momentum
and the rapidity distributions of the photon
as well as the distribution of  the
azimuthal angle - rapidity distance $R_{b \gamma}$
between the  photon and the  hardest  $b$-jet.
Finally, we show the NLO QCD $K$-factor as a
function of photon transverse momentum where factorization and renormalization scales are set to $\mu = m_t$.
}
\label{fig3}
\end{figure}

\begin{figure}[t]
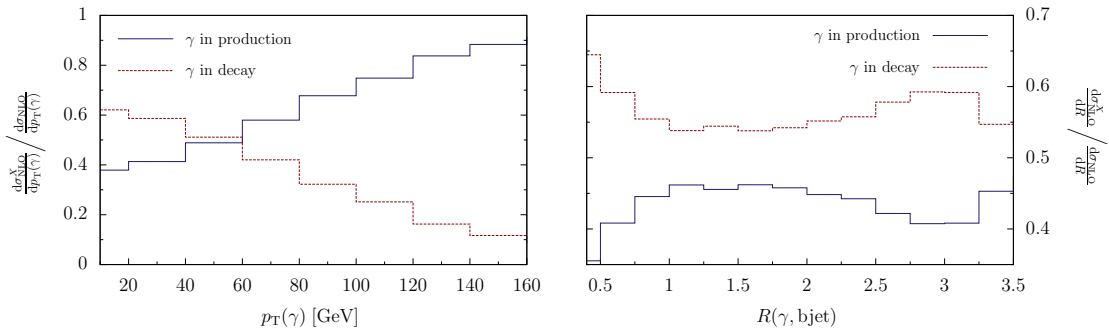

 \begin{center}
  \scalebox{0.5}{\input{TEV_24_Fig09.tex}}
  \scalebox{0.5}{\input{TEV_24_Fig10.tex}}
\end{center}
\caption{
Fraction of events originating
from photon radiation in the production  $p \bar p \to t t \gamma$ and
decay $ t \to W  b  \gamma$ processes,
computed at NLO QCD.
The renormalization and factorization scales are set equal to
$\mu = m_t$.
}
\label{fig5}
\end{figure}

\subsection{$t\bar t \gamma$ production at the Tevatron}
\label{tev_dec}
The CDF collaboration has recently measured the production
cross-section of the $p \bar p \to t \bar t \gamma$ process \cite{cdf}. A
new analysis is under way \cite{privatecom}  that will extend
it to a larger data sample.  It is therefore of interest
to compute the
NLO QCD corrections to
$p \bar p \to
(t \to W^+(l^+\nu) b  )
(\bar t \to W^- (jj) \bar b ) \gamma $ using selection criteria
that are employed in the ongoing  analysis.
Following Refs.~\cite{cdf,privatecom}, we
impose constraints  on transverse momenta and rapidities
of leptons, photons and jets in the process
\ba
&& p_{\perp, \ell} > 20~{\rm GeV},\;\;\;
p_{\perp, \gamma} >  10~{\rm GeV},\;\;\;p_{\perp,j} > 15~{\rm GeV};\\
\nonumber
&& |y_{\ell}| < 1.1,\;\;\;
\;\;\;\;|y_\gamma| < 1.1,\;\;\;\;|y_{j}| < 2.
\ea
In addition we require  that there is missing transverse energy in the event
$E_{\perp, \rm miss} > 20~{\rm GeV}$ and that the transverse energy
$H_\perp$ is larger than $200~{\rm GeV}$. We define the transverse
energy as  $H_\perp = E_{\perp,\rm miss} +
\sum_{i} E_{\perp,i}$, where the sum includes the charged lepton, the photon
 and  jets in the event.
The photon must be
isolated  from jets   $R_{\gamma j} =\sqrt{(y_\gamma - y_j)^2
+ (\varphi_\gamma - \varphi_j)^2} > 0.4$
and  leptons
$R_{\gamma l} =\sqrt{(y_\gamma - y_l)^2
+ (\varphi_\gamma - \varphi_l)^2} > 0.4$.
The photon isolation is implemented following~Ref.~\cite{Frixione:1998jh};
we described it in Section~\ref{sect2}.
We define jets using  the $k_\perp$-clustering
algorithm  \cite{cat,esop}
with $\Delta R = 0.4$, and
require that  at least three
jets are reconstructed, two of which are  $b$-jets.
The $b$-jets are defined as
jets that contain $b$-quarks from top decays, when partons
are clustered according to the jet algorithm.

With the cuts specified above,  we obtain
the following results for the total cross-sections at leading and next-to-leading
order in QCD
\be
\sigma_{\rm LO} = 2.85^{+1.14}_{-0.75}~{\rm fb},\;\;\;
\sigma_{\rm NLO} = 2.64^{+0.21}_{-0.03}~{\rm fb}.
\label{eq456}
\ee
The central values correspond to the renormalization and factorization scales
set to $\mu = m_t$ and lower(upper) value to $\mu = 2 m_t$($m_t/2$),
respectively.  The scale uncertainty in the NLO cross-section is
reduced by a factor
of four, compared to the leading-order one.
For $\mu = m_t$, the NLO QCD corrections
reduce the cross-section by about eight percent,
similar to stable top quark results
discussed in Section~\ref{sect2}.

\begin{figure}[t]
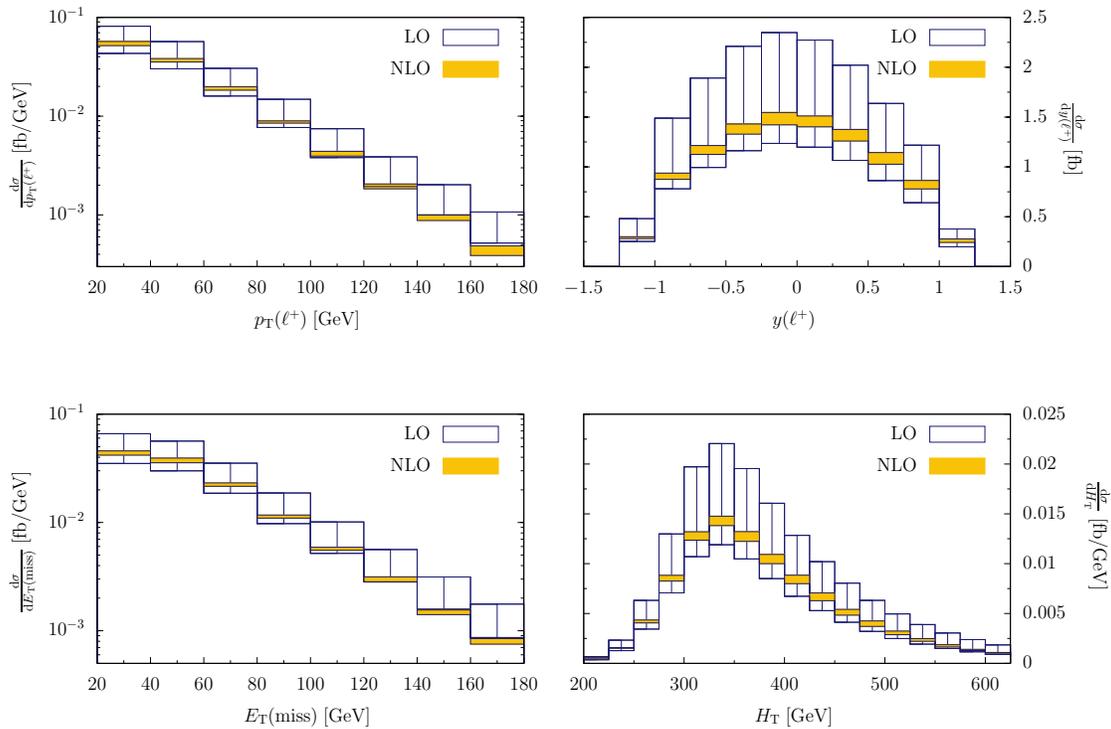

 \begin{center}
  \scalebox{0.5}{\input{TEV_24_Fig03.tex}}
  \scalebox{0.5}{\input{TEV_24_Fig04.tex}} \\[8mm]
  \scalebox{0.5}{\input{TEV_24_Fig05.tex}}
  \scalebox{0.5}{\input{TEV_24_Fig06.tex}} 
\end{center}
\caption{
Kinematic distributions
in $p \bar p \to
(t \to W^+(l^+\nu) b  )
(\bar t \to W^-(jj) \bar b ) \gamma $ process
at the Tevatron. The bands correspond to the variation of
the renormalization and factorization scales
in the interval $m_t/2 < \mu <  2m_t$.
We show the transverse momentum
and the rapidity distributions
of the charged lepton
and distributions of the missing energy and $H_\perp$.
}
\label{fig4}
\end{figure}

In Fig.~\ref{fig3} we show kinematic distributions of
photons in the process
$p \bar p \to
(t \to W^+(l^+\nu) b  )
(\bar t \to W^- (jj) \bar b ) \gamma $.  We observe a significant reduction
in the scale dependence for all kinematic distributions.
The shapes of these distributions do not change much although
the photon transverse momentum distribution
 becomes somewhat softer.
 This is illustrated in the lower-right pane in Fig.~\ref{fig3} where
the local $K$-factor $K = {\rm d} \sigma_{\rm NLO}/{\rm d} \sigma_{\rm LO}$
is shown in dependence of the
photon transverse momentum.
Assuming that  the integrated
luminosity of $10$ inverse femtobarns will, eventually, be analyzed
by the Tevatron collaborations, $t \bar t$ pairs accompanied
by  photons with transverse momenta
as high as $\sim 100~{\rm GeV}$ should be observable.
In Fig.~\ref{fig5} we separately show fractions
of accepted events  where the photon is radiated either in
 the production or in the  decay process,
computed through NLO in perturbative QCD.  We observe that
low-$p_\perp$ photons are produced with comparable
probabilities in the $t \bar t $
production
and decay stages, while photons with
high transverse momentum $p_{\perp,\gamma}  > 80~{\rm GeV}$ are mostly radiated in the production
stage. In Figure~\ref{fig4}, we show
distributions of the charged lepton transverse momentum
and rapidity, as well as distributions of missing energy and $H_{\rm T}$.
Distributions of lepton transverse momentum and  missing transverse
energies become softer at next-to-leading order.
The reduction
in the forward-backward asymmetry for top quarks, discussed in
the previous Section,
is visible in the rapidity distribution of the charged lepton.

Finally, we remark that
the cross-sections shown
in Eq.(\ref{eq456}) correspond to our best approximation
to the setup of the experimental analysis\footnote{The implementation
of photon isolation
in the experimental analysis is different from what we use in this
paper.} described in Ref.~\cite{cdf}, and it is
interesting to estimate the number of $t \bar t \gamma$ events  that
this cross-section value corresponds to.
To do so, we take the NLO QCD cross-section
shown in Eq.(\ref{eq456}), multiply it by a factor of four, to account
for the possibility to produce $e^\pm, \mu^\pm$ final states and
multiply by a factor of $0.22$ which roughly reflects the
experimental efficiencies \cite{privatecom}.
Finally, we multiply by the luminosity $1.9~{\rm fb}^{-1}$ which corresponds
to the data sample analyzed in Ref.~\cite{cdf} and obtain $N_{\rm events}
= 4 \times 0.22 \times 1.9~{\rm fb}^{-1} \times 2.64~{\rm fb} \approx 4$.
It is peculiar that in the experimental analysis \cite{cdf} nine $t \bar t \gamma$ events were observed.
Since a measurement that uses a larger data sample is under way, it will be interesting to see what happens to this difference.
Repeating the above estimate for a luminosity of $6~{\rm fb}^{-1}$, we find 14 $t \bar t \gamma$ events.


\subsection{$t \bar t \gamma$ production at the LHC}

Next, we study the
$p p \to
(t \to W^+(l^+\nu) b  )
(\bar t \to W^- (jj) \bar b ) \gamma $
process  at the LHC with the  center of mass energy of
14~TeV.  We apply the following generic cuts
that describe  detector acceptances and the trigger
\ba
&&
p_{\perp,\gamma} > 20~{\rm GeV},\;\;\;
|y_\gamma| < 2.5,\;\;\;\;
R_{\gamma,b}  > 0.4,\;\;\;
R_{\gamma,j}  > 0.4,\;\;\;
R_{\gamma,\ell} > 0.4,
\nonumber \\ &&
p_{\perp,b} > 20~{\rm GeV},\;\;\;
p_{\perp,j} > 20~{\rm GeV},\;\;\;
p_{\perp,\ell} > 20~{\rm GeV},\;\;\;
E_{\perp,\mathrm{miss}} > 20~{\rm GeV},
\nonumber \\ &&
|y_b| < 2.0,\;\;\;
|y_j| < 2.5,\;\;\;
|y_\ell| < 2.5.
\label{eq_bc}
\ea
We require that there are two $b$-jets and at least
two light jets in the
event.  Jets  are defined using the $k_\perp$-clustering algorithm
\cite{cat,esop} with $\Delta R=0.4$.
We require large transverse energy $H_\perp > 200~{\rm GeV}$.
The photon isolation is implemented following Ref.~\cite{Frixione:1998jh}.
Using these  cuts we obtain the cross-sections
for  $pp \to t \bar t \gamma$ production
\be
\sigma_{\rm LO} = 74.50^{+23.98}_{-16.89}~{\rm fb},\;\;\;
\sigma_{\rm NLO} = 138^{+30}_{-23}~{\rm fb}.
\label{eq70}
\ee
The central values correspond to the renormalization and factorization scales
set to $\mu = m_t$ and lower(upper) value to $\mu = 2 m_t$($m_t/2$),
respectively.
It follows from Eq.(\ref{eq70}) that QCD corrections are rather large.
We saw in Section~\ref{sect2} that NLO
QCD corrections to $pp \to t \bar t \gamma$
process,  evaluated in the approximation of stable top
quarks, increase
the production cross-section by a factor $1.5$,
for $\mu = m_t$. It follows from Eq.(\ref{eq70}) that when
acceptance cuts Eq.(\ref{eq_bc}) are applied, the $K$-factor
increases to $1.86$ and there is only marginal decrease in the
scale dependence.
We have checked that this increase
is related to the radiation of an
additional hard jet in $pp \to t \bar t \gamma$
process and that this enhancement disappears when the
additional jet is  required
to be relatively soft. The large scale dependence of the NLO prediction
for the cross-section is caused by the contribution of the quark-gluon partonic
annihilation channel that only appears at next-to-leading order.
While this effect exists at both the Tevatron and the LHC, it gets significant
enhancement at the LHC due to a much larger gluon luminosity.
We note that similar $K$-factors and large residual scale dependence
can also be observed  in $pp \to t \bar t$ production at the LHC,
when basic kinematic cuts
are applied to
top quark decay products.

\begin{figure}[!t]
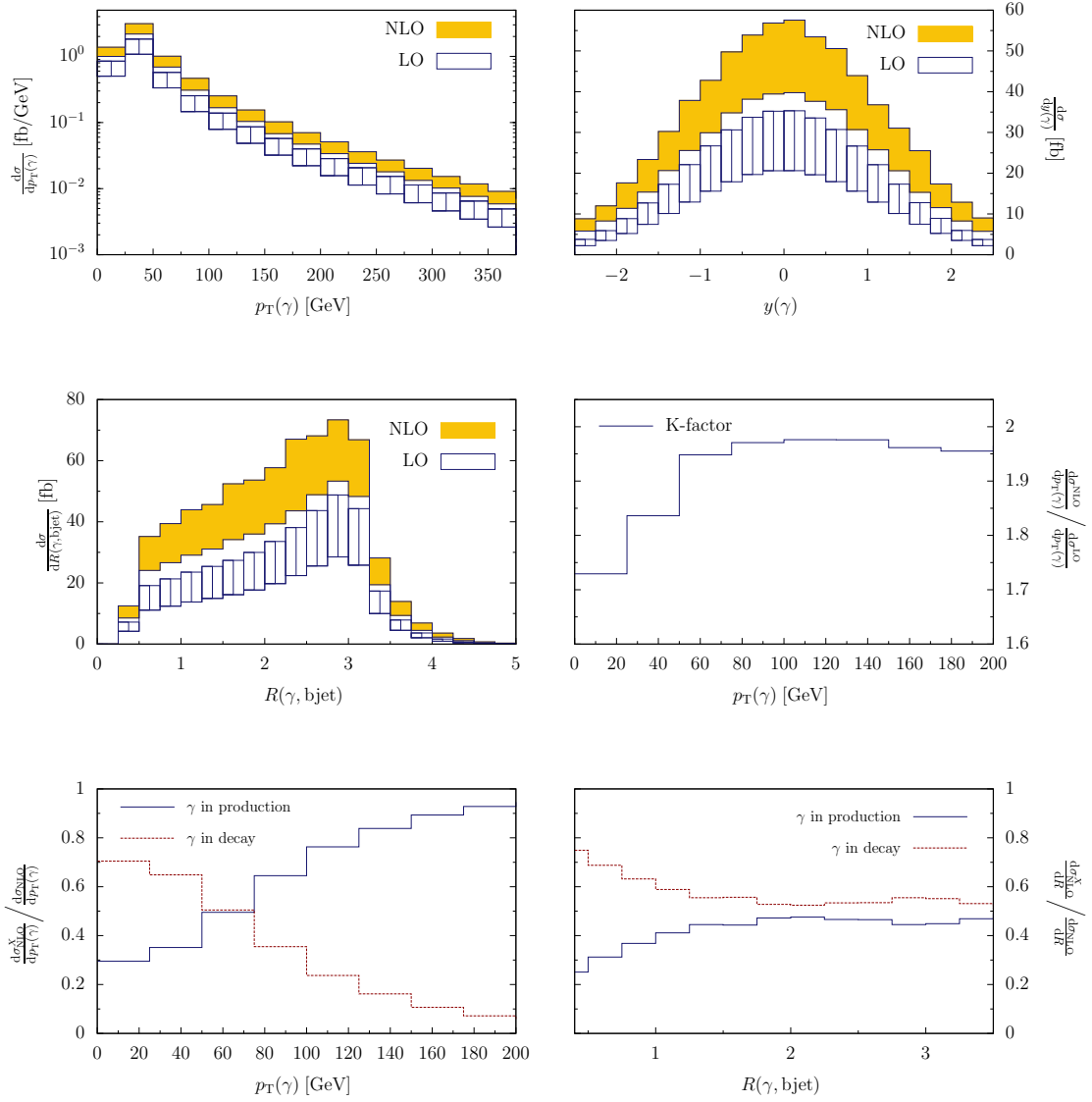

 \begin{center}
  \scalebox{0.5}{\input{LHC_28_Fig01.tex}}
  \scalebox{0.5}{\input{LHC_28_Fig02.tex}} \\[8mm]
  \scalebox{0.5}{\input{LHC_28_Fig07.tex}}
  \scalebox{0.5}{\input{LHC_28_Fig11.tex}}\\[8mm]
  \scalebox{0.5}{\input{LHC_28_Fig09.tex}}
  \scalebox{0.5}{\input{LHC_28_Fig10.tex}}
\end{center}
\caption{
Kinematic distributions
in $p  p \to
(t \to W^+(l^+\nu) b  )
(\bar t \to W^- (jj) \bar b ) \gamma $ process
at the $14$~TeV
LHC, using cuts specified
in Eq.(\ref{eq_bc}).
The bands correspond to the variation of
the renormalization and factorization scales
in the interval $m_t/2 < \mu <  2m_t$.
We show distributions of the transverse momentum and the
rapidity of the photon, as well
as the distribution of  the rapidity-azimuth distance
between the photon and the hardest $b$-jet. We also
show the $K$-factor in dependence on the photon transverse momentum
and the fraction of events for photon radiation in the $t \bar t$ production
and $t(\bar t)$ decay stage.
}
\label{fig6}
\end{figure}

\begin{figure}[!t]
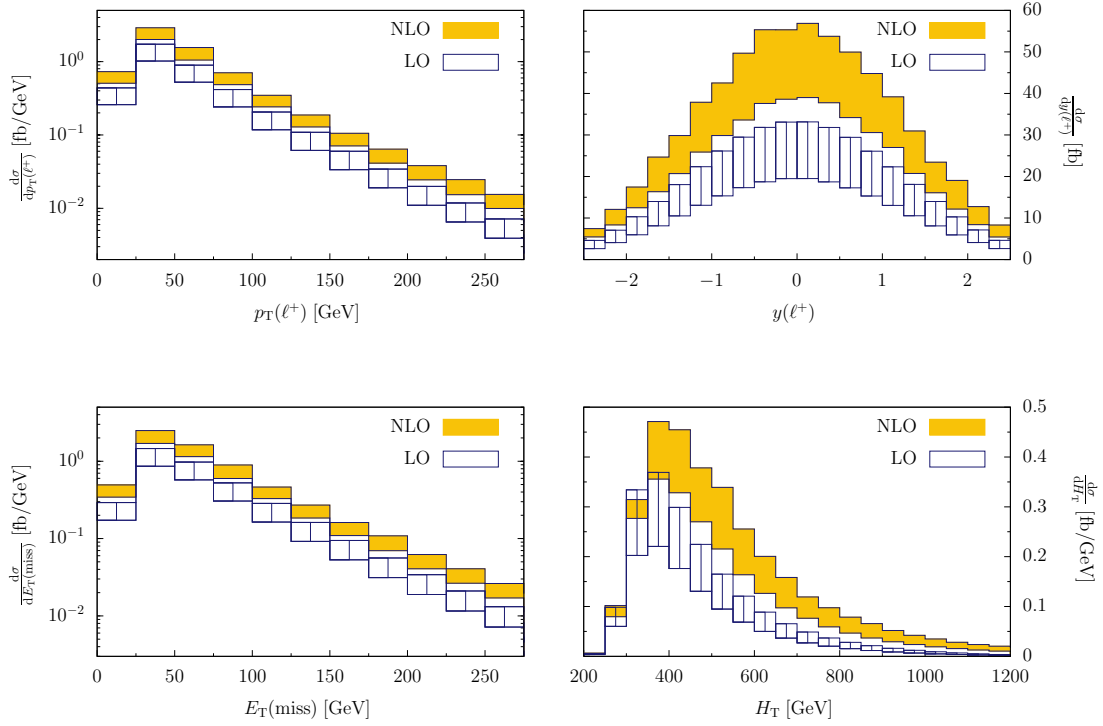

 \begin{center}
  \scalebox{0.5}{\input{LHC_28_Fig03.tex}}
  \scalebox{0.5}{\input{LHC_28_Fig04.tex}} \\[8mm]
  \scalebox{0.5}{\input{LHC_28_Fig05.tex}}
  \scalebox{0.5}{\input{LHC_28_Fig06.tex}} 
\end{center}
\caption{
Kinematic distributions
in $p p \to
(t \to W^+(l^+\nu) b  )
(\bar t \to W^- (jj) \bar b ) \gamma $
at the $14~{\rm TeV}$ LHC, using cuts specified
in Eq.(\ref{eq_bc}).
The bands correspond to the variation of
the renormalization and factorization scales
in the interval $m_t/2 < \mu <  2m_t$.
We show transverse momentum and rapidity distributions
of the charged lepton as well as distributions of
the missing transverse energy $E_{\perp,\rm  miss}$ and
the transverse energy $H_\perp$.
}
\label{fig7}
\end{figure}

In Figs.~\ref{fig6},\ref{fig7} we show various kinematic distributions
at the LHC. Among other things, we observe a dominance of the contribution
from photon radiation in top quark decays over radiation in the $t \bar t$
pair production, for $p_{\perp,\gamma} < 60~{\rm GeV}$.   As an illustration,
we quote results for the NLO QCD cross-sections where photon
radiation occurs either in the production or in the decay stage
\be
\sigma_{\rm prod}^{\rm NLO} = 60.9~{\rm fb},\;\;\;
\sigma_{\rm dec}^{\rm NLO} =  77.2~{\rm fb}.
\ee
These results correspond to the factorization and the renormalization
scales set to the top quark mass, $\mu = m_t$; their sum gives the total NLO
cross-section shown  in Eq.(\ref{eq70}).
We also see that the spectrum of
emitted photons becomes {\it harder}, in contrast to the Tevatron
case.  The $K$-factor, as a function of the photon transverse momentum
is shown in Fig.~\ref{fig6}.  From Fig.~\ref{fig6} we estimate that
photons with transverse momenta of up to $350~{\rm GeV}$ should be
observable at the $14~{\rm TeV}$ LHC with ten inverse femtobarns
of accumulated luminosity.

\begin{figure}[!t]
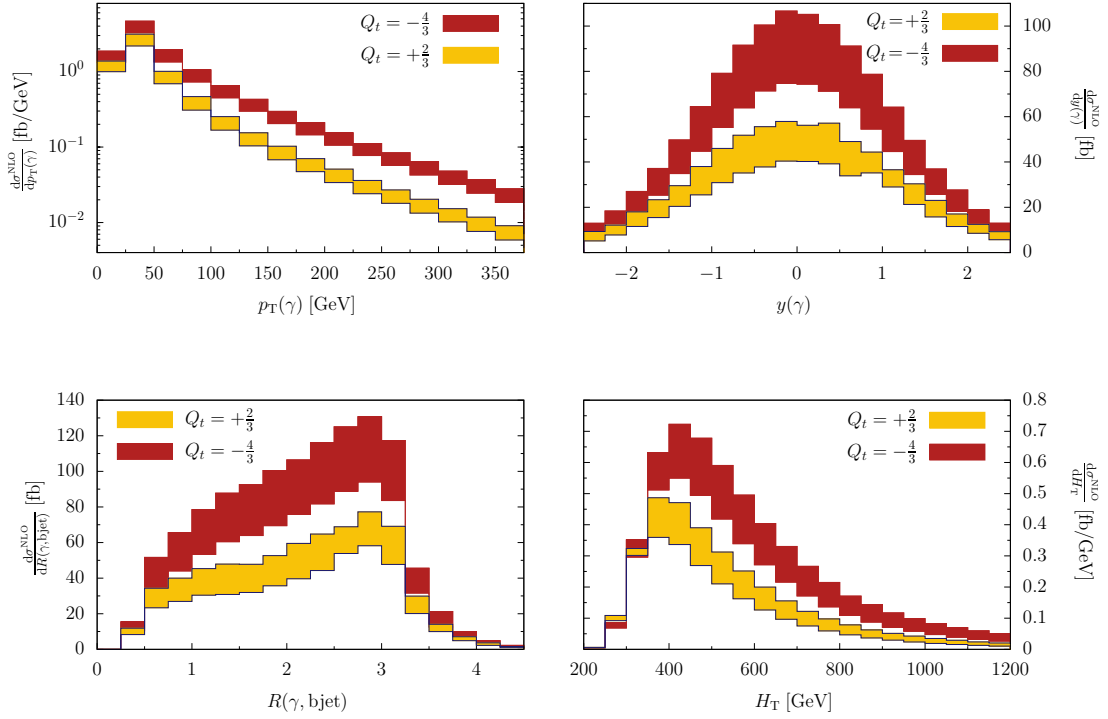

 \begin{center}
  \scalebox{0.5}{\input{LHC_28XQ_Fig01.tex}}
  \scalebox{0.5}{\input{LHC_28XQ_Fig02.tex}} \\[8mm]
  \scalebox{0.5}{\input{LHC_28XQ_Fig07.tex}}
  \scalebox{0.5}{\input{LHC_28XQ_Fig06.tex}}
\end{center}
\caption{
Kinematic distributions
in $p  p \to t \bar t (\gamma) \to
l^+ \nu b \bar b jj + \gamma$
at the $14$~TeV LHC for two top quark electric charges.
The bands correspond to the variation of
the renormalization and factorization scales
in the interval $m_t/2 < \mu <  2m_t$ at NLO QCD.
}
\label{fig8}
\end{figure}

As mentioned earlier, the production of $t \bar t \gamma$ at the LHC
can be used to constrain electromagnetic couplings of the top quark
and, in particular, its electric charge. In what follows we compare some
results for top quarks with charges $Q_t = 2/3$ and $Q_t = -4/3$.
Needless to say that the ``top'' quark with the charge $Q_t = -4/3$
decays through $t \to b W^-$, to respect  electric charge
conservation.   With acceptance cuts
shown in Eq.(\ref{eq_bc}), we find
\be
\sigma_{\rm LO}^{Q_t = -4/3} =136.8^{+46.7}_{-32.3};\;\;\;
\sigma_{\rm NLO}^{Q_t = -4/3} =243^{+50}_{-39}~{\rm fb}.
\label{qt43xs}
\ee
The central values correspond to the renormalization and factorization scales
set to $\mu = m_t$ and lower(upper) value to $\mu = 2 m_t$($m_t/2$),
respectively.
These results, together with the cross-sections values
shown in Eq.(\ref{eq70}),
imply that the ratio of $Q_t = -4/3$ and $Q_t = 2/3$
cross-sections is
\be
{\cal R}^{\rm LO}
= \frac{\sigma_{\rm LO}^{Q_t = -4/3}}{\sigma_{\rm LO}^{Q_t = 2/3}}
 = 1.84^{+0.02}_{-0.03},\;\;\;
{\cal R}^{\rm NLO}
= \frac{\sigma_{\rm NLO}^{Q_t = -4/3}}{\sigma_{\rm NLO}^{Q_t = 2/3}}
 = 1.76^{+0.01}_{-0.02}.
\label{eqratio}
\ee
The uncertainty in the ratio comes from the dependence of the production
cross-sections for the two top quark charges on the renormalization
and the factorization scales $\mu$. In Fig.~\ref{fig8} we compare basic
kinematic distributions for two top quark charge
assignments, $Q_t = 2/3$ and $Q_t = -4/3$.

We note that the dominance of $gg$ annihilation at the LHC seems
to suggest that $\sigma(pp \to t \bar t \gamma)$ cross-sections
should scale like the electric charge of the top quark to second power,
so that the naive expectation for ${\cal R}$ in Eq.(\ref{eqratio})
is ${\cal R} = 4$. It is obvious from Eq.(\ref{eqratio}) that this
expectation fails.  This happens because photons are dominantly
radiated by the decay products of $t$ and $\bar t$ (cf. Fig.~\ref{fig6})
 and this contribution  does not scale as $Q_t^2$.
It is also interesting to remark that  shifts from leading to next-to-leading
order and the remaining NLO uncertainties  in the cross-section
ratios shown in Eq.(\ref{eqratio}) are quite small, in particular when
compared to the corresponding uncertainties in the cross-sections,
Eqs.(\ref{eq70},\ref{qt43xs}).  In fact, it is easy to imagine  that large
changes in $pp \to t \bar t \gamma$
cross-sections, from leading to next-to-leading order,
are not particular to  $t \bar t \gamma $ production and originate, rather,
from the underlying dynamics of $pp \to t \bar t$ process.  To prove that
this assertion is valid, we compute ratios of
$p  p \to t \bar t (\gamma) \to
l^+ \nu b \bar b jj + \gamma$
to $p  p \to t \bar t \to
l^+ \nu b \bar b jj$  cross-sections,
subject to basics cuts
shown in Eq.(\ref{eq_bc}),
at the $14~{\rm TeV}$ LHC. We  obtain
\be
\frac{\sigma_{t \bar t \gamma}^{Q_t = 2/3}}
{\sigma_{t \bar t} }
= \left \{
\begin{array}{cc}
 5.66^{+0.03}_{-0.02} \times 10^{-3},  & {\rm LO}; \\
 6.33^{+0.26}_{-0.14} \times 10^{-3}, & {\rm NLO},
\end{array}
\right.
\;\;\;\;
\frac{\sigma_{t \bar t \gamma}^{Q_t = -4/3}}{\sigma_{t \bar t} }
= \left \{
\begin{array}{cc}
 10.4^{+0.2}_{-0.2} \times 10^{-3},  & {\rm LO};\\
 11.2^{+0.3}_{-0.2} \times 10^{-3}, & {\rm NLO}.
\end{array}
\right.
\ee

\begin{figure}[t]
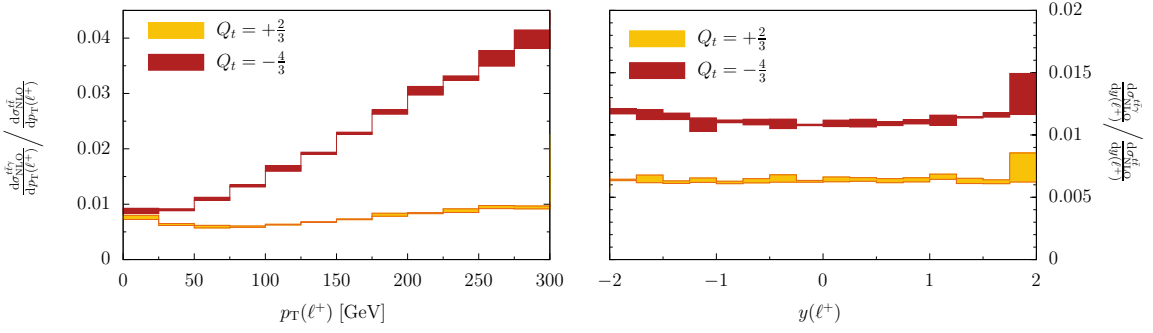

 \begin{center}
  \scalebox{0.5}{\input{LHC_28Norm_Fig01.tex}}
  \scalebox{0.5}{\input{LHC_28Norm_Fig02.tex}}
\end{center}
\caption{
Transverse momentum and rapidity distribution of the charged lepton
for two values of the top quark electric charge in
$p  p \to t \bar t (\gamma) \to l^+ \nu b \bar b jj + \gamma$
at the $14~{\rm TeV}$ LHC,
normalized to $p  p \to t \bar t  \to l^+ \nu b \bar b jj$.
In the latter case the top quark
charge is kept at its canonical
value  $Q_t = 2/3$.
}
\label{fig11}
\end{figure}

It is clear that ratios of cross-sections are significantly  more stable
against radiative corrections and
scale variations than the cross-sections themselves.
Moreover, these ratios help reduce scale uncertainties
in  kinematic distributions as well.
We illustrate this in Fig.~\ref{fig11} where we show lepton
 kinematic distributions in  $pp \to t \bar t \gamma$ for $Q_t =2/3$
and $Q_t = -4/3$
at next-to-leading order, normalized
to similar distributions in $pp \to t \bar t$.  In both cases, basic cuts
shown in Eq.(\ref{eq_bc}) are applied. It is striking that
for $Q_t = 2/3$, the ratio
${\rm d}\sigma_{t \bar t \gamma}/{\rm d}\sigma_{t \bar t}$
is essentially constant for a large range of kinematic parameters,
while for $Q_t =-4/3$ the relevant spectra appear to be harder.
However, in both cases the scale uncertainty of the ratio is much smaller
than the scale uncertainty when  $pp \to t \bar t \gamma$ {\it and}
$pp \to t \bar t$ are considered separately. It is clear that,
in addition to the scale uncertainty, other uncertainties such as in
$\alpha_s$ and in parton distribution functions cancel to a large extent
in the cross-section ratio \cite{Bernreuther:2008ju}, making it
an interesting observable to study at the LHC.

Although the ratio of $Q_t = -4/3$ and $Q_t = 2/3$ cross-sections
shown
in Eq.(\ref{eqratio}) appears already large enough to distinguish
between the two electric charge assignments, one can make it even larger.
Indeed, as we already mentioned,
at the LHC  top quarks are mostly produced in gluon collisions. Hence,
in the production stage of the process,
photons are radiated mostly  by top quarks. If we manage to reduce
the contribution from the radiation
in the decay, we will have an observable that is sensitive to the
electromagnetic
couplings of the top quark.
To reduce the probability that the photon is radiated in the top quark
decay, we impose the following cuts:
\begin{itemize}
\item we determine the $b$-jet $j_b$
that forms the smallest invariant mass with the charged lepton
and  require large transverse mass of that  $b$-jet, lepton , photon
and the missing energy
$m_\perp( j_b l \gamma; p_{\perp \rm miss} ) > 180~{\rm GeV}$.
The transverse mass here is defined as
\be
m_\perp^2(j_b l\gamma;  p_{\perp, \rm miss})  =
\left ( \sqrt{p_\perp^2(j_b l \gamma) + m^2(j_b l \gamma) } + p_{\perp, \rm miss}
\right )^2 - (\vec p_\perp(j_bl \gamma) + \vec p_{\perp,\rm miss } )^2;
\ee

\item the remaining $b$-jet is combined with the two hardest light jets;
it is required that the invariant mass of these three jets is close to the top
quark mass
$
160~{\rm GeV}< m(jjj_b) < 180~{\rm GeV};
$

\item  to
suppress photon radiation from leptonic decays of $W$-bosons, we require
$
 m_\perp( \ell \gamma; E_{\perp \rm miss} ) > 90~{\rm GeV};
$

\item to
suppress photon radiation from hadronic decays of $W$-bosons, we require
that there are  two light jets in the event whose invariant mass is close
to the mass of the $W$-boson,
$
70~{\rm GeV} < m(jj) < 90~{\rm GeV}.
$
\end{itemize}

\begin{figure}[t]
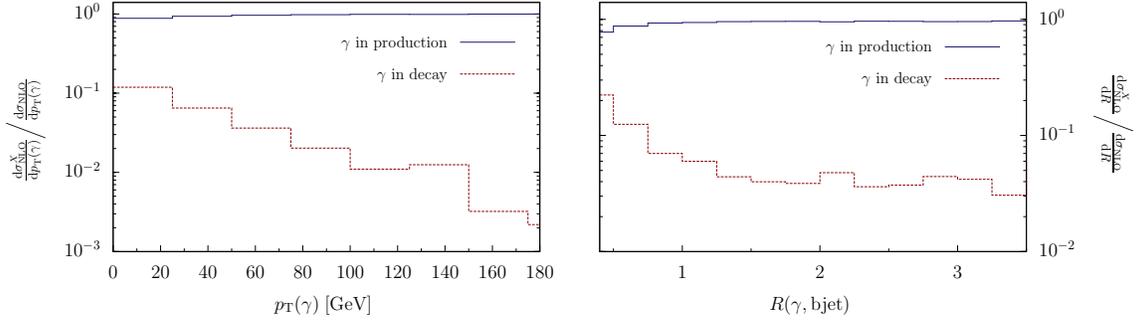

 \begin{center}
  \scalebox{0.5}{\input{LHC_27_Fig09.tex}}
  \scalebox{0.5}{\input{LHC_27_Fig10.tex}}
\end{center}
\caption{
Fractions of $pp \to t \bar t (\gamma) \to l^+\nu b \bar{b} jj+\gamma$ events at the $14~{\rm TeV}$
LHC  with photon radiated in  the production and in the decay, in case
when RDS cuts are applied. The renormalizations scale is
set to $\mu = m_t$.
}
\label{fig9}
\end{figure}

\begin{figure}[t]
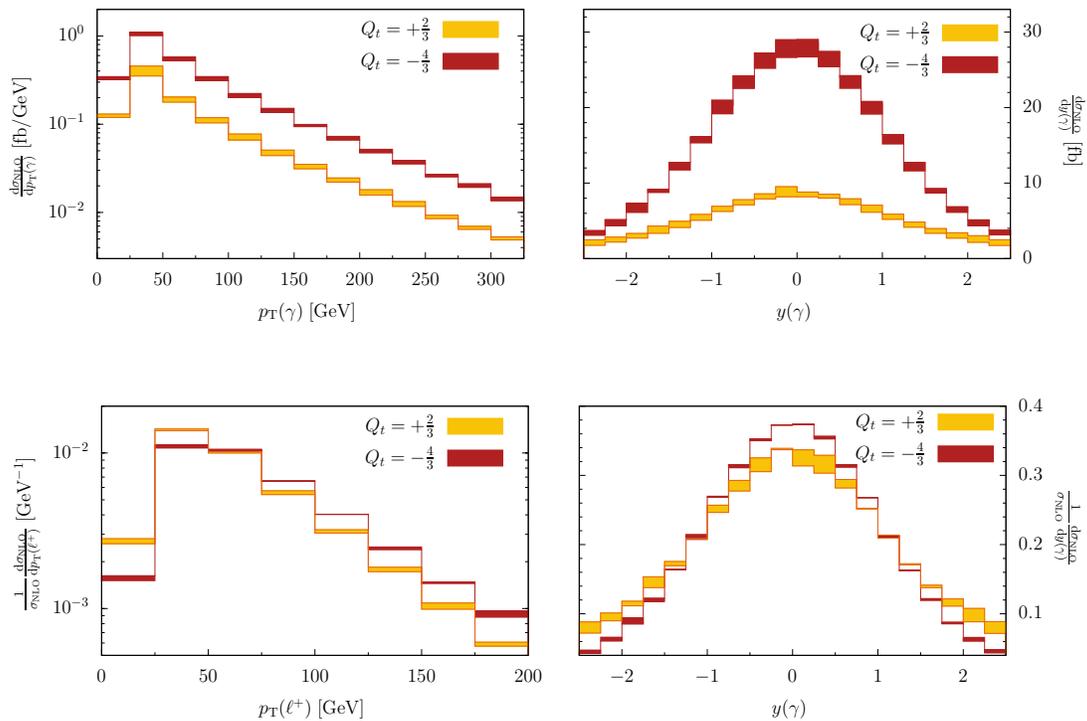

 \begin{center}
  \scalebox{0.5}{\input{LHC_27XQ_Fig01.tex}}
  \scalebox{0.5}{\input{LHC_27XQ_Fig02.tex}} \\[8mm]
  \scalebox{0.5}{\input{LHC_27XQ_Fig04.tex}}%
  \scalebox{0.5}{\input{LHC_27XQ_Fig03.tex}}

\end{center}
\caption{
Kinematic distributions
in $p  p \to t \bar t (\gamma) \to
l^+ \nu b \bar b jj + \gamma$
at the $14~{\rm TeV}$ LHC, for two electromagnetic charge
assignments of top quarks with RDS cuts. The two lower panes show normalized distributions,
to emphasize differences in shapes.
The bands correspond to the variation of
the renormalization and factorization scales
in the interval $m_t/2 < \mu <  2m_t$.
}
\label{fig10}
\end{figure}

We will refer to those cuts as the ``radiation-in-the-decay-suppression''
(RDS) cuts and we emphasize that they are applied {\it in addition}
to cuts shown in Eq.(\ref{eq_bc}).
Applying RDS cuts, we find that the leading order
$t \bar t \gamma$ production cross-sections
reduces by about a factor of three compared to the case when
only generic
cuts Eq.(\ref{eq_bc}) are applied
\be
\sigma_{\rm LO} = 23.39^{+7.83}_{-5.43}~{\rm fb},\;\;\;
\sigma_{\rm NLO} = 26.7^{+1.3}_{-2.3}~{\rm fb}.
\label{qt23rds}
\ee
Compared to cross-sections shown in Eq.(\ref{eq70}), there are
significant changes in the NLO result as well since the
$K$-factor becomes much smaller when  RDS cuts are applied.
Kinematic distributions are shown in Figs.~\ref{fig9},\ref{fig10}.
The degree of suppression for photons originating from top quark
decays can be seen in Fig.~\ref{fig9} where the distribution
of the photon transverse momentum and the
angular distance between the photon and the hardest $b$-jet
$R_{\gamma, j_b}$ are displayed.
We observe that with the RDS cuts
more than ninety percent of the total cross-section is due to photons
radiated in the production of a $t \bar t$ pair with about five percent
coming from top quark decays.

Finally,
we apply the RDS cuts to  compute the cross-section for the production
of the top quark with the exotic charge $Q_t=-4/3$. We find
\be
\sigma_{\rm LO}^{Q_t = -4/3} = 72.62^{+25.70}_{-17.61}~{\rm fb},\;\;\;
\sigma_{\rm NLO}^{Q_t = -4/3} =76.9^{+0.5}_{-5.4}~{\rm fb}.
\label{qt43rds}
\ee
where  the central value corresponds  to $\mu = m_t$ and
the lower (upper)  value  to $\mu = 2 m_t$ and $\mu = m_t/2$, respectively.
We determine the ratio  of the cross-sections for the
two charge assignments and find that it increases
\be
{\cal R}^{\rm LO}_{\rm RDS}
= \frac{\sigma_{\rm LO}^{Q_t = -4/3}}{\sigma_{\rm LO}^{Q_t = 2/3}}
 = 3.10^{+0.05}_{-0.04},\;\;\;
{\cal R}^{\rm NLO}_{\rm RDS}
= \frac{\sigma_{\rm NLO}^{Q_t = -4/3}}{\sigma_{\rm NLO}^{Q_t = 2/3}}
 = 2.88^{+0.05}_{-0.12},
\label{eqratio1}
\ee
compared to Eq.(\ref{eqratio}) where only basic cuts
are applied.

We can now estimate if it is worth applying the RDS cuts.
We denote by ${\cal L}$ the luminosity required
to separate $Q_t=-4/3$ from $Q_t = 2/3$ at the $3\sigma$ level
with the cuts in Eq.(\ref{eq_bc}) and by
${\cal L}_{\rm RDS}$ the same quantity when the RDS cuts are applied
in addition.
The two quantities are related by the following equation\footnote{We only
consider statistical errors.}
\be
\frac{{\cal L}}{{\cal L}_{\rm RDS}} =
\frac{\sigma^{Q_t = 2/3}_{\rm RDS}}{\sigma^{Q_t = 2/3}}
\frac{({\cal R}_{\rm RDS}-1)^2}{({\cal R} -1 )^2}
\ee
We can use Eqs.(\ref{eq70},\ref{qt43xs},\ref{eqratio},\ref{qt23rds},\ref{qt43rds},\ref{eqratio1}) to compute the
ratio of the required luminosities at leading and next-to-leading
order in perturbative QCD. Interestingly, because the $K$-factors
for the two types of cuts are so different, we find that
the required ratios of luminosities differ by a significant amount
\be
\frac{{\cal L}}{{\cal L}_{\rm RDS}} =
\left \{
\begin{array}{cc}
1.98 \pm 0.02,  & {\rm LO}; \\
1.12 \pm 0.08, & {\rm NLO}.
\end{array}
\right.
\label{eq_lumi}
\ee
It follows from Eq.(\ref{eq_lumi}) that once  next-to-leading order
effects are accounted for,
the application of  RDS suppression cuts becomes much less important
since a factor of two gain in luminosity gets reduced to ${\cal O}(10\%)$
gain. We also find that  kinematic distributions are not very sensitive
to the top quark charge; for illustrative purposes
we show some distributions that have some sensitivity to
$Q_t$ in Fig.~\ref{fig10}. The largest effect
is present in the lepton transverse momentum distribution
that becomes harder when the top quark charge  increases.
Also, the rapidity distribution of the charged lepton becomes
more central, but this effect is not very significant.


\section{Conclusions}
\label{sectc}
In this paper, we describe the calculation of the NLO QCD corrections to the
production of a $t \bar t$ pair in association with
a hard photon at the Tevatron and the LHC. This process is of interest for
direct studies of the electromagnetic couplings of the top quark such as its
electric charge and its anomalous magnetic moment.
In a recent measurement, the CDF collaboration detected
nine $t\bar{t}\gamma$ events in $p \bar{p}$ collisions
using an integrated luminosity of $1.9\;\mathrm{fb}^{-1}$.
Our best estimate, which includes
NLO QCD corrections, realistic acceptances
and  photon radiation in the production and decay stages of a
$t \bar t$ pair, is that $4.4 \pm 0.2$  events should have been observed  with
the integrated luminosity of $1.9~{\rm fb}^{-1}$ and an efficiency of $22\,\%$.
It will be interesting to compare our results with an analysis of a
larger data sample, that is under way \cite{privatecom}.

For  any hadron collider process, an accurate prediction requires
at least next-to-leading order QCD computation and this is what
we set out to do in this paper for the $t \bar t \gamma$ final state.
For stable top quarks
such a computation was performed in Ref.~\cite{ma}. However, for practical
studies of $t \bar t \gamma$ production, the result of Ref.~\cite{ma}
is not sufficient since, in general, large fraction of isolated photons
comes from radiative decays of top quarks. It is important
to incorporate top quark decays, including radiative ones, into a
unified  framework that also includes  higher-order QCD corrections,  and
this is what we accomplished in this paper.

We studied the $p \bar p (pp) \to t \bar t (\gamma) \to
l^+ \nu b \bar b jj + \gamma
$ process both
at the Tevatron and at the $14~{\rm TeV}$ LHC, including effects
of the NLO QCD corrections.
We found that, in general, the  QCD corrections are small at the
Tevatron and are large at the LHC. However, we also observed that these
QCD corrections  are very similar to the NLO QCD corrections to
$p \bar p (pp) \to t \bar t$ processes suggesting that
ratios of these cross-sections
${\rm d} \sigma_{t \bar t \gamma}/{\rm d} \sigma_{t \bar t}$
can be theoretically predicted with higher accuracy than the two cross-sections
separately.

We found that about fifty percent of all  photons in
the $t \bar t \gamma$
events  are  radiated off the
top quark decay products, both at the Tevatron and the LHC.
This fraction increases with the decrease in the photon transverse
momentum  reaching approximately eighty percent for
$p_{\perp, \gamma} \sim 10~{\rm GeV}$
at the LHC.
Since photon emission off the top quark decay products is a background
to measuring electromagnetic couplings of the top quarks, it is important
to apply selection criteria that suppresses these contributions.
Designing cuts to suppress photon  radiation in top quark decays
and applying them  to $pp \to t \bar t \gamma$ process at the LHC,
we observe a significant change in the $K$-factor compared to
the $K$-factor computed for basic cuts. This feature emphasizes the
importance of flexible implementation of the radiative corrections
to processes with unstable particles, where kinematics of  the decay
products must  be accessible.

To have a concrete model where the electric charge  of the top
quark is different from its Standard Model value, we have studied
the case when $Q_t = -4/3$ and the ``top quark'' decays into
the  $b$-quark and the $W^-$-boson \cite{ma,ub}.
We have used this charge assignment to investigate the possibility
of measuring $Q_t$ by studying $t \bar t \gamma$ production at the
LHC. If the analysis is performed at LO QCD, we find that
designing  cuts  to suppress the
 QCD radiation from top quark decays benefits the analysis.
 However, when the same analysis is performed at NLO QCD,
suppressing  QCD radiation off the decay products of top quarks
becomes
less important  because, when basic cuts are applied,  the
$t \bar t \gamma$
production process receives large corrections
at  next-to-leading order in perturbative QCD.

Finally, we note that the production of photons in association
with a $t \bar t$ pair can be studied at the $7$ TeV LHC. For the
selection criteria as in Eq.(\ref{eq_bc}),  we find
the leading and next-to-leading order cross-section
for $pp \to t \bar t (\gamma) \to
l^+ \nu b \bar b jj  + \gamma$
to be $15~{\rm fb}$ and $26~{\rm fb}$, respectively.
If $5~{\rm fb}^{-1}$  are indeed collected at the $7~{\rm TeV}$
LHC by the end of the year 2012, we estimate that
about $500$ events with high energy isolated photons, large missing energy,
an isolated lepton and two $b$-jets should be observed at the LHC.  Among these
events, there will  be a few truly spectacular ones,
with  the $t \bar t$ pair accompanied by a very energetic photon.  We look
forward to studies of the $pp \to t \bar t \gamma$ process  at the LHC
in the coming years.\\


{\bf Acknowledgments}
We would like to thank   Benjamin Auerbach, Henry Frisch
and Avto Kharchilava  for useful communications.
We greatly benefited from conversations with U. Baur (deceased).
This research  is supported by the NSF under grants
PHY-0855365 and PHY-0547564,
and by the startup
funds provided by Johns Hopkins University.
Calculations reported in this paper were performed on the Homewood
High Performance Cluster of Johns Hopkins University.

\begin{appendix}

\section{ Gauge-invariant decomposition of scattering amplitudes
with a photon
}
\label{a1}

In this Appendix, we present the
gauge invariant decomposition of
helicity amplitudes that we
used in the  calculation reported in this paper.
Our starting point is the computation of the $t \bar t j$  hadroproduction process
reported in Ref.~\cite{msj}. From that reference, we know primitive
amplitudes \cite{bdk} for partonic processes such as
$ 0 \to t \bar t ggg$,
$0 \to t \bar t q \bar q g$ etc. and we would like
to turn them into
amplitudes that describe production of a $t \bar t$ pair,
quarks, gluons and a single
photon.
We do so by constructing  linear combinations of $t\bar{t} + {\rm gluons} + {\rm quarks}$
 primitive amplitudes in such a way that non-Abelian contributions cancel.
Because the photon can also be radiated in
the decay of the top quark,
amplitudes for $t \bar t$ pair production without photon radiation
are also required.
Those amplitudes can be found e.g. in Ref.~\cite{ms}, but we present them
here for completeness as a special case of
the $t\bar{t}\gamma$ amplitudes.

We begin with the amplitudes for
leading order processes for $t\bar{t}\gamma$ production.
At leading order, two partonic initial states $gg$ and $q \bar{q}$
contribute to the cross section.
We write the  color decomposition of the corresponding matrix elements in the following
form
\begin{eqnarray}
    \mathcal{M}^{\mathrm{tree}}(gg\rightarrow t\bar{t}\gamma)
    &=&
    g_s^2 \; \sqrt{2}\, e Q_t \;
   \sum_{\sigma \in S_2} (T^{a_{\sigma_3}} T^{a_{\sigma_4}})_{i_2}^{\bar{i}_1} \; \mathcal{A}^{\mathrm{tree}} (1_{\bar t},5_g, 2_t,(\sigma_3)_g,(\sigma_4)_g),
\label{eqn:tree1}
\\
    \mathcal{M}^{\mathrm{tree}}(q\bar{q}\rightarrow t\bar{t}\gamma)
    &=&
    g_s^2 \; \sqrt{2}\, e Q_t \; \Big[
   \delta^{\bar{i}_1}_{i_4} \delta^{\bar{i}_3}_{i_2}
  -\frac{1}{N_\mathrm{c}} \delta^{\bar{i}_1}_{i_2} \delta^{\bar{i}_3}_{i_4}
   \Big]  \mathcal{B}^{\mathrm{tree}} (1_{\bar t},2_t,3_{\bar{q}},4_q;5_\gamma).
\label{eqn:tree1a}
\end{eqnarray}
In Eq.(\ref{eqn:tree1a}) we use
\begin{eqnarray}
    \mathcal{B}^{\mathrm{tree}} (1_{\bar t},2_t,3_{\bar{q}},4_q;5_\gamma)
    =
      \mathcal{A}^{\mathrm{tree}} (1_{\bar t},5_g,2_t,3_{\bar{q}},4_q)
    + \frac{Q_q}{Q_t} \,
\mathcal{A}^{\mathrm{tree}} (1_{\bar t},2_t,3_{\bar{q}},5_g,4_q).
\label{eqn:tree2}
\end{eqnarray}
We note that tree  partial amplitudes
$\mathcal{A}^{\mathrm{tree}}$ in
Eqs.(\ref{eqn:tree1},\ref{eqn:tree2})
involve only quarks and gluons, i.e. no photon.
The factor $\sqrt{2}$ in Eq.(\ref{eqn:tree1},\ref{eqn:tree1a})
compensates for the similar factor in the color-stripped  Feynman rules
\cite{rules}
that
we use to compute amplitudes with quarks and gluons. Because
we require photon, rather than gluon,
 emission amplitudes, we must remove this factor.

To obtain amplitudes for the $t\bar{t}$ final state, without photon radiation,
we use the following set of rules:
\begin{itemize}
\item  remove $ 5_g$ from partial and primitive amplitudes;
\item set $\sqrt{2}\, e Q_t \rightarrow 1$  and $ Q_q, Q_f\rightarrow 0$;
\item  set auxiliary parameter $\kappa \rightarrow 0$.
\end{itemize}
These rules can be applied to
Eqs.(\ref{eqn:tree1},\ref{eqn:tree1a},\ref{eqn:tree2}) as well as to all
other  photon-emission amplitudes that we present  in this Section.

We now describe amplitudes required for the next-to-leading order QCD computation.
For the real emission corrections we require  four partonic channels
$ gg \rightarrow t \bar{t}  \gamma g,\; q \bar{q} \rightarrow t \bar{t}  \gamma g,\; qg \rightarrow t \bar{t}  \gamma q$ and $\bar{q}g \rightarrow t \bar{t} \gamma \bar{q}$. The
last three channels  are related by crossing symmetry. For this reason,
we present the color decomposition for  $gg  \to t \bar t g \gamma $ and
$q \bar q \to t \bar t \gamma g$. We find
\begin{eqnarray}
    \mathcal{M}^{\mathrm{real}}(gg\rightarrow t\bar{t} \gamma g)
    &=&
    g_s^3 \; \sqrt{2}\, e Q_t \;
    \sum_{\sigma \in S_3} (T^{a_{\sigma_3}} T^{a_{\sigma_4}} T^{a_{\sigma_6}} )_{i_2}^{\bar{i}_1} \; \mathcal{A}^{\mathrm{tree}} (1_{\bar t},5_g, 2_t,(\sigma_3)_g,(\sigma_4)_g,(\sigma_6)_g),\;
\\
    \mathcal{M}^{\mathrm{real}}(q\bar{q}\rightarrow t\bar{t}\gamma g)
    &=&
    g_s^3 \; \sqrt{2}\, e Q_t \; \bigg[
        (T^{a_6})^{\bar i_1}_{i_4} \delta^{\bar i_3}_{i_2} \;\mathcal{B}^{\mathrm{tree}}_1 (1_{\bar t},2_t,3_{\bar{q}},4_q,6_g;5_\gamma)
\nonumber \\ && \quad\quad\quad\quad
       +\; (T^{a_6})^{\bar i_3}_{i_2} \delta^{\bar i_1}_{i_4} \; \mathcal{B}^{\mathrm{tree}}_2 (1_{\bar t},2_t,3_{\bar{q}},4_q,6_g;5_\gamma)
\nonumber\\ && \quad\quad\quad\quad
       +\; \frac1{N_\mathrm{c}} (T^{a_6})^{\bar i_1}_{i_2} \delta^{\bar i_3}_{i_4}  \; \mathcal{B}^{\mathrm{tree}}_3 (1_{\bar t},2_t,3_{\bar{q}},4_q,6_g;5_\gamma)
\nonumber \\ && \quad\quad\quad\quad
       +\; \frac1{N_\mathrm{c}} (T^{a_6})^{\bar i_3}_{i_4} \delta^{\bar i_1}_{i_2}  \; \mathcal{B}^{\mathrm{tree}}_4 (1_{\bar t},2_t,3_{\bar{q}},4_q,6_g;5_\gamma)
   \bigg],
\end{eqnarray}
where
\begin{eqnarray}
    \mathcal{B}^{\mathrm{tree}}_1 (1_{\bar t},2_t,3_{\bar{q}},4_q,6_g;5_\gamma)
    &=&
      \mathcal{A}^{\mathrm{tree}} (1_{\bar t},5_g,2_t,3_{\bar{q}},4_q,6_g)
    + \frac{Q_q}{Q_t} \, \mathcal{A}^{\mathrm{tree}} (1_{\bar t},2_t,3_{\bar{q}},5_g,4_q,6_g),
\\
    \mathcal{B}^{\mathrm{tree}}_2 (1_{\bar t},2_t,3_{\bar{q}},4_q,6_g;5_\gamma)
    &=&
      \mathcal{A}^{\mathrm{tree}} (1_{\bar t},5_g,2_t,6_g,3_{\bar{q}},4_q)
    + \frac{Q_q}{Q_t} \, \mathcal{A}^{\mathrm{tree}} (1_{\bar t},2_t,6_g,3_{\bar{q}},5_g,4_q),
\\
    \mathcal{B}^{\mathrm{tree}}_3 (1_{\bar t},2_t,3_{\bar{q}},4_q,6_g;5_\gamma)
    &=&
    \frac{\kappa+1}{2} \bigg(  \mathcal{A}^{\mathrm{tree}} (1_{\bar t},5_g,6_g,2_t,3_{\bar{q}},4_q)
    + \mathcal{A}^{\mathrm{tree}} (1_{\bar t},6_g,5_g,2_t,3_{\bar{q}},4_q)
\nonumber \\ &&
 + \frac{Q_q}{Q_t} \, \mathcal{A}^{\mathrm{tree}}(1_{\bar t},6_g,2_t,3_{\bar{q}},5_g,4_q) \bigg),
\\
    \mathcal{B}^{\mathrm{tree}}_4 (1_{\bar t},2_t,3_{\bar{q}},4_q,6_g;5_\gamma)
    &=&
      \mathcal{A}^{\mathrm{tree}}(1_{\bar t},5_g,2_t,3_{\bar{q}},6_g,4_q)
    + \frac{Q_q}{Q_t} \,\bigg( \mathcal{A}^{\mathrm{tree}} (1_{\bar t},2_t,3_{\bar{q}},5_g,6_g,4_q)
\nonumber \\ &&
    +  \mathcal{A}^{\mathrm{tree}} (1_{\bar t},2_t,3_{\bar{q}},6_g,5_g,4_q) \bigg).
\end{eqnarray}
As long as we are interested in $t \bar t \gamma$ amplitudes, $\kappa$ should be set to one.
For $t \bar t$ amplitudes without photon emission, $\kappa$ should be set to zero.

The matrix elements of the virtual corrections are constructed from primitive amplitudes.
For the partonic channel $gg\rightarrow t\bar{t} \gamma$ we find the color decomposition
\begin{eqnarray}
    \mathcal{M}^{\mathrm{virt}}(gg\rightarrow t\bar{t} \gamma)
    =
    g_s^4 \; \sqrt{2}\, e Q_t  \sum_{i=\mathrm{a,b}}
   \sum_{\sigma \in S_2}    \bigg[(T^{a_{\sigma_3}} T^{a_{\sigma_4}})_{i_2}^{\bar{i}_1}
   \mathcal{B}^{\mathrm{virt}}_{1,i} (1_{\bar t}, 2_t,(\sigma_3)_g,(\sigma_4)_g;5_\gamma) &
\nonumber \\
   +
   \mathrm{Tr}(T^{a_{\sigma_3}} T^{a_{\sigma_4}}) \, \delta^{\bar i_1}_{i_2}  \mathcal{B}^{\mathrm{virt}}_{2,i} (1_{\bar t}, 2_t,(\sigma_3)_g,(\sigma_4)_g;5_\gamma)
   \bigg] &
\end{eqnarray}
with
\begin{eqnarray}
    \mathcal{B}^{\mathrm{virt}}_{1,\mathrm{a}} (1_{\bar t}, 2_t,3_g,4_g;5_\gamma)
    &=&
    N_\mathrm{c}\, \mathcal{A}^{\mathrm{L},[1]} (1_{\bar t},5_g, 2_t,3_g,4_g)
   -\frac{2\kappa+1}{3 N_\mathrm{c}} \bigg(
   \mathcal{A}^{\mathrm{L},[1]} (1_{\bar t},4_g,5_g, 3_g,2_t)
\nonumber\\ &&
    + \mathcal{A}^{\mathrm{L},[1]} (1_{\bar t},5_g,4_g, 3_g,2_t) + \mathcal{A}^{\mathrm{L},[1]} (1_{\bar t},4_g,3_g, 5_g,2_t)
   \bigg),
\\
    \mathcal{B}^{\mathrm{virt}}_{1,\mathrm{b}} (1_{\bar t}, 2_t,3_g,4_g;5_\gamma)
    &=&
   \sum_f \bigg\{
   \mathcal{A}^{\mathrm{L},[1/2]}_f(1_{\bar t},5_g,2_t,3_g,4_g)
   +\frac{Q_f}{Q_t} \bigg(
    \mathcal{A}^{\mathrm{L},[1/2]}_f(1_{\bar t},2_t,3_g,4_g,5_g)
\nonumber\\ &&
    +\mathcal{A}^{\mathrm{L},[1/2]}_f(1_{\bar t},2_t,3_g,5_g,4_g)
    +\mathcal{A}^{\mathrm{L},[1/2]}_f(1_{\bar t},2_t,5_g,3_g,4_g)
\nonumber\\ &&
    +\mathcal{A}^{\mathrm{L},[1/2]}_f(1_{\bar t},5_g,2_t,3_g,4_g)
    \bigg) \bigg\}, \label{eqn:fsum1}
\\
    \mathcal{B}^{\mathrm{virt}}_{2,\mathrm{a}} (1_{\bar t}, 2_t,3_g,4_g;5_\gamma)
    &=&
    \mathcal{A}^{\mathrm{L},[1]} (1_{\bar t},5_g,2_t,3_g,4_g)
    + \frac{2\kappa+1}{3} \bigg(\mathcal{A}^{\mathrm{L},[1]} (1_{\bar t},5_g,4_g,3_g,2_t)
\nonumber\\ &&
    + \mathcal{A}^{\mathrm{L},[1]} (1_{\bar t},4_g,5_g,3_g,2_t) +  \mathcal{A}^{\mathrm{L},[1]} (1_{\bar t},4_g,3_g,5_g,2_t) \bigg)
\\ &&
    + \frac{\kappa+1}{2} \bigg( \mathcal{A}^{\mathrm{L},[1]} (1_{\bar t},5_g,3_g,2_t,4_g) +  \mathcal{A}^{\mathrm{L},[1]} (1_{\bar t},3_g,5_g,2_t,4_g) \bigg)
     , \quad \nonumber
\\
\label{eqn:fsum2}
    \mathcal{B}^{\mathrm{virt}}_{2,\mathrm{b}} (1_{\bar t}, 2_t,3_g,4_g;5_\gamma)
    &=&
    -\frac1{N_\mathrm{c}} \mathcal{B}^{\mathrm{virt}}_{1,\mathrm{b}} (1_{\bar t}, 2_t,3_g,4_g;5_\gamma),
\end{eqnarray}
\begin{figure}[t]
\label{fig:primamps}
\begin{center}
\epsfig{file=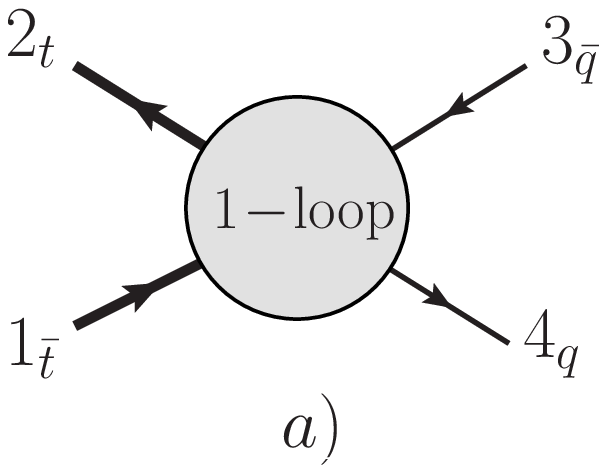, angle=0,width=0.2\textwidth}
\hspace{1cm}
\epsfig{file=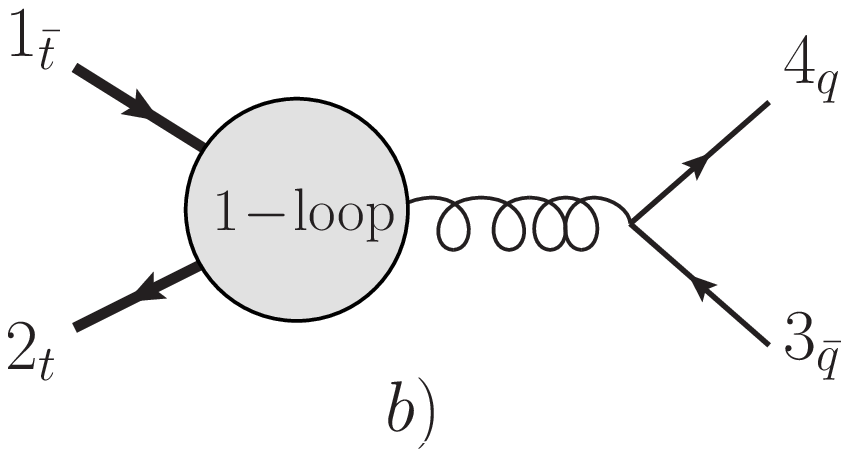, angle=0,width=0.27\textwidth}
\\[0.3cm]
\epsfig{file=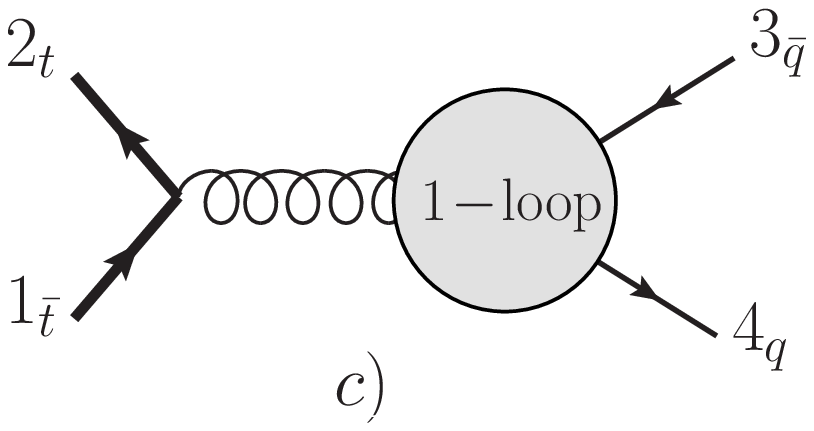, angle=0,width=0.27\textwidth}
\hspace{1cm}
\epsfig{file=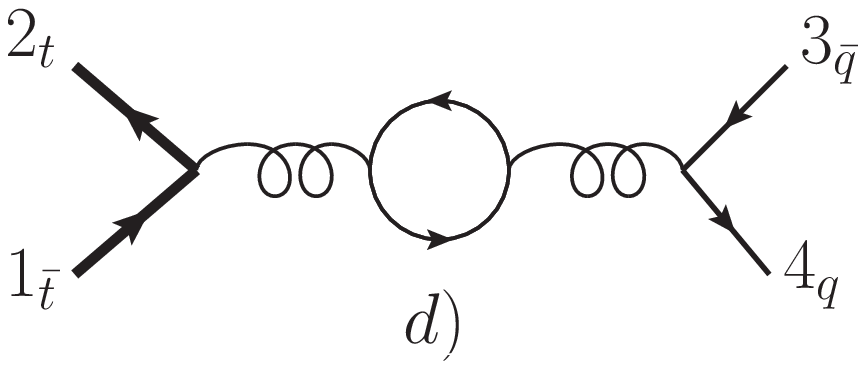, angle=0,width=0.29\textwidth}
\caption{Definition of primitive amplitudes with four quarks.
      Figure~(a) defines
 $\mathcal{A}_{a}$ which includes all topologies where both
external fermion lines enter the loop.
    Figures~(b) and (c) define $\mathcal{A}_{b}$
  and $\mathcal{A}_{c}$ where either the top quark
   or the light quark line enter the loop, respectively.
   Finally, Figure~(d) defines amplitude $\mathcal{A}_{d}$
   that includes topologies with a closed fermion loop.}
\end{center}
\end{figure}
where $\mathcal{A}^{\mathrm{L},[1]}$
and $\mathcal{A}^{\mathrm{L},[1/2]}_f$ are regular primitive amplitudes as defined in \cite{bdk}.
The sum over $f$ in Eq.(\ref{eqn:fsum1}) and Eq.(\ref{eqn:fsum2}) includes all quark flavors.
The color decomposition of the matrix element for the process $q\bar{q} \rightarrow t\bar{t} \gamma$
follows Ref.~\cite{egkmz} and is given by
\begin{eqnarray}
    \mathcal{M}^{\mathrm{virt}}(q\bar{q}\rightarrow t\bar{t} \gamma)
    =
    g_s^4 \; \sqrt{2}\, e Q_t \!\! \sum_{i=\mathrm{a,b,c,d}}   &\bigg[ &
    \delta^{\bar i_1}_{i_4} \delta^{\bar i_3}_{i_2} \mathcal{B}^\mathrm{virt}_{1,i}(1_{\bar t}, 2_t,3_{\bar q},4_q;5_\gamma)
\nonumber \\
   &
+&\delta^{\bar i_1}_{i_2} \delta^{\bar i_3}_{i_4} \mathcal{B}^\mathrm{virt}_{2,i}(1_{\bar t}, 2_t,3_{\bar q},4_q;5_\gamma)
   \bigg]
\end{eqnarray}
with
\begin{eqnarray}
    \mathcal{B}^\mathrm{virt}_{1,\mathrm{a}}(1_{\bar t}, 2_t,3_{\bar q},4_q;5_\gamma)
    &=&
    \bigg(N_\mathrm{c}-\frac2{N_\mathrm{c}}\bigg) \bigg( \mathcal{A}_\mathrm{a}(1_{\bar t},5_g, 2_t,3_{\bar q},4_q)
    + \frac{Q_q}{Q_t} \mathcal{A}_\mathrm{a}(1_{\bar t}, 2_t,3_{\bar q},5_g,4_q)  \bigg)
\nonumber\\ &&
    -\frac2{N_\mathrm{c}}\bigg( \mathcal{A}_\mathrm{a}(1_{\bar t},5_g, 2_t,4_q,3_{\bar q})
    - \frac{Q_q}{Q_t} \mathcal{A}_\mathrm{a}(1_{\bar t}, 2_t,4_q,5_g,3_{\bar q})   \bigg),
\\
    \mathcal{B}^\mathrm{virt}_{1,\mathrm{b}}(1_{\bar t}, 2_t,3_{\bar q},4_q;5_\gamma)
    &=&
    -\frac{2\kappa+1}{3 N_\mathrm{c}}\bigg(
    \mathcal{A}_\mathrm{b}(1_{\bar t},5_g, 4_q,3_{\bar q},2_t)
   +\mathcal{A}_\mathrm{b}(1_{\bar t}, 4_q,5_g,3_{\bar q},2_t)
\nonumber\\&&
   +\mathcal{A}_\mathrm{b}(1_{\bar t}, 4_q,3_{\bar q},5_g,2_t)
   - \frac{Q_q}{Q_t} \mathcal{A}_\mathrm{b}(1_{\bar t}, 4_q,5_g,3_{\bar q},2_t) \bigg),
\\
    \mathcal{B}^\mathrm{virt}_{1,\mathrm{c}}(1_{\bar t}, 2_t,3_{\bar q},4_q;5_\gamma)
    &=&
    -\frac1{N_\mathrm{c}}\bigg(
    \mathcal{A}_\mathrm{c}(1_{\bar t},5_g,2_t,3_{\bar q},4_q)
    - \frac{Q_q}{Q_t}
    \bigg\{  \mathcal{A}_\mathrm{c}(1_{\bar t},2_t,5_g,3_{\bar q},4_q)
\nonumber\\&&
    + \mathcal{A}_\mathrm{c}(1_{\bar t},2_t,3_{\bar q},4_q,5_g)
    + \mathcal{A}_\mathrm{c}(1_{\bar t},5_g,2_t,3_{\bar q},4_q)
    \bigg\}
    \bigg),
\\
    \mathcal{B}^\mathrm{virt}_{1,\mathrm{d}}(1_{\bar t}, 2_t,3_{\bar q},4_q;5_\gamma)
    &=&
    \sum_f \bigg( \mathcal{A}_\mathrm{d}^f(1_{\bar t},5_g,2_t,3_{\bar q},4_q)
   +\frac{Q_q}{Q_t} \mathcal{A}_\mathrm{d}^f(1_{\bar t},2_t,3_{\bar q},5_g,4_q) \bigg).
\end{eqnarray}
The primitive amplitudes $\mathcal{A}_{\mathrm{a,b,c,d}}$ correspond
to different topologies depending on
which fermion lines enter the loop, see Fig.~13. 
We note that the contribution of the amplitude
$B_{2,i}^{\rm virt}$ vanishes after its interference with the tree
amplitude is computed; for this reason, we do not present it
here.

\section{ Formula for the production cross-section }

\newcommand{\tb}{\bar{t}}
\newcommand{\rd}{\mathrm{d}}
\newcommand{\LO}{\mathrm{LO}}
\newcommand{\NLO}{\mathrm{NLO}}
\newcommand{\dsL}{\; \mathrm{d}\sigma^\mathrm{LO}}
\newcommand{\dsN}{\; \mathrm{d}\sigma^{\delta \mathrm{NLO}}}
\newcommand{\dGL}{\; \mathrm{d}{\cal B}^\mathrm{LO}}
\newcommand{\dGN}{\; \mathrm{d}{\cal B}^{\delta \mathrm{NLO}}}

In this Appendix, we present the formula that we use to describe
radiative corrections
to top quark pair production
in association with a photon  in the narrow width approximation.
In the narrow width approximation, the differential production cross-section is given
by
\be
{\rm d}\sigma  =
{\rm d} \sigma_{t \bar t \gamma} {\rm d} {\cal B}_{t,X} {\rm d} {\cal B}_{\bar t,Y}
+ {\rm d} \sigma_{t \bar t} \left [ {\rm d} {\cal B}_{t,X\gamma}
{\rm d} {\cal B}_{\bar t,Y}
+ {\rm d} {\cal B}_{t,X} {\rm d} {\cal B}_{\bar t,Y \gamma} \right ],
\ee
where  ${\cal B}_{t,X(\gamma)}$ is the branching fraction for
 either radiative $t \to X + \gamma$
or non-radiative $t \to X$
top quark decay. The above
equation is valid to all orders in the strong coupling
constant. We expand through  ${\cal O}(\alpha_s^3)$ and denote
\be
{\cal B}_{i,X}^{\rm LO} =
\frac{{\rm d} \Gamma^{\rm LO}_{i \to X}}{\Gamma^{\rm LO}_t},\;\;\;
{\cal B}_{i,X}^{\delta \rm NLO} =
\frac{{\rm d} \Gamma^{\delta \rm NLO}_{i \to X}}{\Gamma^{\rm LO}_t},\;\;\;
\;\;\; \chi = 1 - 2
\frac{\Gamma_{t}^{\delta \rm NLO}}{\Gamma_{t}^{\rm LO}},
\ee
where $i \in [t, \bar t]$, $\Gamma^{\rm LO}_t$ is the total
decay width of the top quark at leading order and
$\Gamma^{\delta \rm LO}_t$ is the QCD correction to the total
top quark decay width.
We obtain
\begin{eqnarray}
\label{eqn:masterDK}
{\rm d} \sigma
&=&
    \chi \dsL_{t\tb} \left( \dGL_{\tb, Y \gamma} \dGL_{t,X}
     + \dGL_{\tb, Y} \dGL_{t, X\gamma} \right)+
    \chi \dsL_{t\tb\gamma} \dGL_{\tb,Y} \dGL_{t,X}
\nonumber \\
    & + &
    \dsN_{t\tb} \left( \dGL_{\tb, Y \gamma}\dGL_{t,X}
 + \dGL_{\tb, Y}\dGL_{t,X\gamma} \right)
\nonumber \\
    & + &
    \dsL_{t\tb} \Big( \dGN_{\tb, Y \gamma} \dGL_{t,X} + \dGL_{\tb,Y}
    \dGN_{t, X \gamma}
+  \dGN_{\tb,Y} \dGL_{t, X\gamma} + \dGL_{\tb,Y \gamma} \dGN_{t, X}  \Big)
\nonumber \\
    & + &
    \dsN_{t\tb\gamma} \dGL_{\tb, Y} \dGL_{t, X}
     +
    \dsL_{t\tb\gamma} \left( \dGN_{\tb, Y} \dGL_{t, X}
    + \dGL_{\tb, Y} \dGN_{t, X} \right)
+{\cal O}(\alpha_s^4).
\end{eqnarray}
Since we also treat the $W$-boson in the narrow width approximation,
photon radiation in the top quark decay can be further decomposed  into
photon radiation in top quark  ($t \to Wb$) and $W$ ($W \to f \bar f'$) decays.  The
decays of the $W$-bosons are treated at  leading order in perturbative
QCD. We therefore write
\begin{eqnarray}
    {\rm d} {\cal B}_{t,X \gamma}^i =
\frac{\rd\Gamma_{bW\gamma}^i}{\Gamma^{\LO}_t}
\; \frac{\rd\tilde\Gamma_{W \to f\bar{f'}}^{\text{LO}}}{\Gamma_{W}}
+ \frac{\rd\Gamma_{bW}^i}{{\Gamma^{\LO}_t} }
\; \frac{\rd\tilde\Gamma_{W \to f\bar{f'}\gamma}^{\text{LO}}}{\Gamma_{W}},
\;\;\;\;i \in [ {\rm LO}, \delta {\rm NLO} ],\;\; X = b f \bar f'.
\end{eqnarray}


\end{appendix}

\end{document}